\documentclass[fleqn,10pt]{wlscirep}
\usepackage[utf8]{inputenc}
\usepackage[T1]{fontenc}

\usepackage[version=4]{mhchem}

\usepackage{lineno}
\usepackage{amsmath}
\usepackage{subcaption}
\usepackage{bm}
\usepackage{graphicx}

\usepackage[format=plain,justification=justified,singlelinecheck=false,font={stretch=1.125,small,sf},labelfont=bf,labelsep=space]{caption}

\newcommand{\ie}[0]{\textit{i.e.}, }

\usepackage{chemformula}
\usepackage{braket}
\newcommand{\DefineAuthor}[2]{%
  \expandafter\newcommand\csname #1note\endcsname[1]{%
    \textbf{\textcolor{#2}{\textbf{#1:} ##1}}}%
  \expandafter\newcommand\csname #1\endcsname[1]{
    \textbf{\textcolor{#2}{##1}}}
  \expandafter\newcommand\csname #1cancel\endcsname[1]{%
    \textbf{\textcolor{#2}{\sout{##1}}}}%
  \expandafter\newcommand\csname #1change\endcsname[2]{%
    \textbf{\textcolor{#2}{\sout{##1} ##2}}}%
  \newenvironment{#1text}{\color{#2}}{\color{black}}
}

\definecolor{dartmouthgreen}{rgb}{0.05, 0.5, 0.06}
\DefineAuthor{LMS}{blue}

\usepackage{xcolor}
\usepackage[normalem]{ulem}

\title{QALPA: Property-guided diffusion modeling for efficient exploration of chemical spaces of flexible molecules}

\author[1,$\dag$]{Michael Hanna}
\author[2,$\dag$]{Julian Cremer}
\author[1,3,4]{Zekiye Erarslan}
\author[3,4,5,*]{Leonardo Medrano Sandonas}

\affil[1]{Faculty of Computer Science, TUD Dresden University of Technology, 01062 Dresden, Germany}
\affil[2]{Machine Learning \& Computational Sciences, Pfizer Worldwide R\&D, Berlin, Germany}
\affil[3]{Center for Advanced Systems Understanding (CASUS), Conrad-Schiedt-Straße 20, Görlitz 02826, Germany}
\affil[4]{Helmholtz Zentrum Dresden-Rossendorf, Bautzner Landstraße 400, Dresden 01328, Germany}
\affil[5]{Institute for Materials Science and Max Bergmann Center of Biomaterials, TUD Dresden University of Technology, 01062 Dresden, Germany}

\affil[$\dag$]{These authors contributed equally to this work.}
\affil[*]{Corresponding author: Leonardo Medrano Sandonas (l.medrano-sandonas@hzdr.de)}

\begin{abstract}
Exploring the chemical space of flexible molecules remains challenging because the vast number of possible compounds and conformations, together with the increasing cost and limited generalization of 3D generative models for larger and more complex molecules, restrict access to unexplored chemistry.
Here, we introduce QALPA (``Quantum-Aware Learning for Property-space Augmentation''), a property-guided generative framework that combines an E(3)-equivariant diffusion model with active learning and efficient quantum-mechanical (QM) methods to iteratively explore targeted QM property manifolds. By coupling generation with physics-based evaluation, QALPA improves molecular sampling and model reliability in sparsely populated regions of chemical space.
Our results show that training on complementary QM datasets spanning both small (QM7-X) and large (Aquamarine) drug-like compounds enables accurate molecular generation across a broad size range, improving transferability beyond the training distribution for complex property manifolds involving both extensive and intensive properties.
As a proof of concept, QALPA coupled with the machine learning-augmented tight-binding method EquiDTB efficiently augments alloQM, a QM dataset introduced in this work, comprising 6,253 conformers of allosteric drug molecules, by populating sparse regions of the property landscape defined by the many-body dispersion energy and HOMO–LUMO energy gap.
These results demonstrate that the integration of generative AI with efficient ML/QM methods offers a practical pathway toward augmenting sparse QM datasets and sustainably expanding the exploration of chemical space for molecular discovery.
\end{abstract}

\begin{document}

\flushbottom
\maketitle

\thispagestyle{empty}

\section{Introduction}

The controlled exploration of the chemical space (CS) of drug-like molecules is a fundamental yet formidable challenge in molecular discovery. The size of this space is estimated to exceed $10^{60}$ compounds, making exhaustive enumeration or screening computationally intractable~\cite{bohacek1996,reymond2010chemicalspace}.
Conventional virtual screening has substantially accelerated early-stage drug discovery by prioritizing promising candidates from existing compound libraries. However, these approaches explore only a minute fraction of the chemically accessible space and remain largely constrained to known molecular scaffolds, leaving vast regions of CS unexplored~\cite{reymond2010chemicalspace,reymond12}.
As a result, the discovery of novel therapeutics continues to be a lengthy and costly process, typically requiring more than a decade of development and investments exceeding two billion dollars per approved drug, while attrition rates remain high throughout the pipeline~\cite{dimasi2016,paul2010nrdd}.
Developing computational strategies capable of efficiently navigating unexplored regions of CS therefore remains a central challenge. Such strategies require informative molecular representations that provide effective ``navigation coordinates'' for identifying promising candidates while avoiding the prohibitive cost of exhaustive exploration\cite{medrano23fod}.

Artificial intelligence (AI) and machine learning (ML) are transforming chemistry and the pharmaceutical pipeline, with applications ranging from mapping chemical space to molecular design~\cite{Tkatchenko2020,VonLilienfeld2020}. Deep learning has enabled remarkable advances, including accurate protein structure and interaction prediction~\cite{jumper2021alphafold,alphafold3}, inverse molecular design that maps target properties to molecular structures~\cite{sanchezlengeling2018,medranosandonas2024inverse}, and the generation of compounds with experimentally validated biological activity~\cite{zhavoronkov2019}.
Among generative approaches, diffusion models---probabilistic methods that learn to reverse a gradual noising process~\cite{ho2020ddpm,song2020sde}---have emerged as particularly effective for molecular and protein design because of their stable training and natural compatibility with three-dimensional (3D) molecular representations.
Representative examples include DiffDock~\cite{corso2023diffdock}, GeoDiff~\cite{xu2022geodiff}, RFdiffusion~\cite{watson2023rfdiffusion}, and equivariant diffusion models (EDM)~\cite{edm}, together with molecular variants that jointly generate atomic coordinates and molecular graphs, such as MiDi~\cite{midi}, EQGAT-diff~\cite{eqgatdiff}, and PILOT~\cite{pilot}.
Collectively, these advances have shifted molecular discovery from screening existing compounds toward \textit{de novo} molecular design, with conditional and property-guided formulations enabling the optimization of target molecular properties during generation~\cite{ho2022classifierfree,pilot}.

This paradigm shift is particularly important as modern drug discovery moves beyond the traditional ``one drug--one target'' framework toward molecules that must simultaneously satisfy multiple, often competing, biological and physicochemical requirements---an inherently multi-property optimization problem~\cite{fromer2023mpo}.
Existing strategies, including evolutionary algorithms~\cite{jensen2019}, reinforcement learning~\cite{olivecrona2017,horwood2020}, and conditional generative models~\cite{pilot}, have demonstrated considerable success but continue to face trade-offs among model expressiveness, architectural complexity, optimization efficiency, and training stability.
These challenges become even more pronounced for flexible molecules, whose high-dimensional conformational landscapes not only substantially increase the computational cost of 3D molecular generation and evaluation, but also decrease output fidelity\cite{susml}.
Furthermore, incorporating accurate quantum-mechanical (QM) calculations into generative workflows to optimize multiple electronic and physicochemical properties dramatically increases computational requirements, limiting the applicability of such approaches to pharmaceutically relevant chemical spaces.
This motivates the integration of hybrid ML/QM frameworks that retain near-QM accuracy while reducing computational costs by several orders of magnitude\cite{wei21,Guoqing22,huang2023central,equidtb}, thereby enabling efficient exploration and optimization of molecular chemical space.

Here, we introduce QALPA (``Quantum-Aware Learning for Property-space Augmentation''), a property-guided diffusion framework for the efficient exploration of chemical spaces populated by flexible molecules. Instead of relying exclusively on explicit structural generation, QALPA navigates a targeted QM property manifold to identify chemically relevant regions while substantially reducing the computational cost of molecular exploration. The framework can operate either as a standalone molecular generator or as a front-end that proposes focused candidate sets for downstream structural refinement. Furthermore, QALPA is embedded within an active learning (AL) workflow that iteratively couples property-guided generation with geometry optimization and QM property evaluation to continuously expand and improve the training data.
QALPA integrates the E(3)-equivariant diffusion model EQDIFF~\cite{eqgatdiff} with large-scale QM datasets spanning both small and drug-like molecules, including QM7-X\cite{qm7x} and Aquamarine\cite{aqm}, to enable efficient learning of molecular property distributions across diverse chemical spaces. In addition, the ML-augmented tight-binding method EquiDTB\cite{equidtb} is incorporated to perform accurate geometry optimization and predict electronic properties of flexible molecules at near DFT-PBE0 accuracy while accounting for many-body dispersion (MBD) interactions.
As a proof of concept, we employ QALPA to augment the sparse QM property landscape of alloQM, a dataset introduced in this work that comprises QM calculations for allosteric drug molecules.
Rather than replacing expert-driven molecular design, QALPA complements it by serving as a high-variance exploration engine capable of efficiently discovering unconventional regions of chemical space, while human expertise and downstream computational workflows provide the low-variance refinement required to identify promising drug candidates.

\section{Methodology}

\subsection{QALPA framework}

\begin{figure}[t!]
    \centering
        \includegraphics[width=0.9\linewidth]{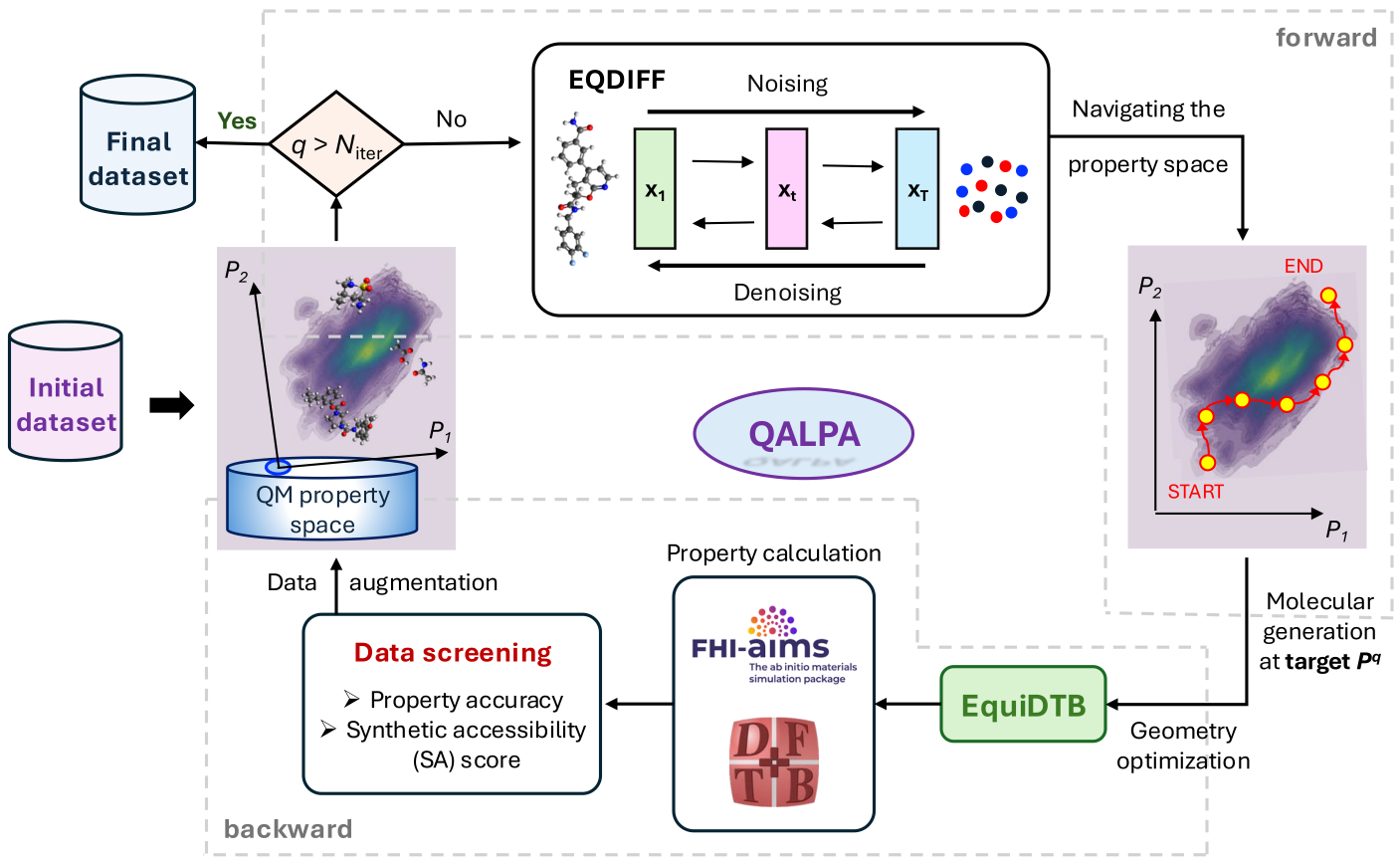}
    \caption{Scheme of the QALPA framework (``Quantum-Aware Learning for Property-space Augmentation''). The forward block trains the EQDIFF generative model on an initial dataset and generates molecules at each navigation coordinate associated with the target property values $P^q={P_1^q,P_2^q}$, where $q$ denotes the active learning (AL) iteration and $P_1$ and $P_2$ define the selected two-dimensional property manifold. The backward block optimizes the generated geometries using EquiDTB (or standard DFTB) and computes the target properties at either the DFT or DFTB level using FHI-aims or DFTB+ packages, respectively. After screening based on property accuracy and synthetic accessibility (SA) score, the selected molecules are added to the training dataset, and the AL loop proceeds until the predefined trajectory has been traversed, \ie $q>N_{\rm iter}$, where $N_{\rm iter}$ is the total number of AL iterations.}
\label{fig1}
\end{figure}

QALPA iteratively augments quantum-mechanical (QM) datasets through an active learning (AL) workflow that alternates between property-guided molecular generation (\textit{forward} process) and physics-based evaluation (\textit{backward} process), see Fig.~\ref{fig1}. Starting from an initial QM dataset, the framework defines a target property manifold that guides the generative model throughout successive AL iterations.
Rather than sampling chemical space indiscriminately, QALPA performs guided exploration using a set of \textit{navigation coordinates}, comprising QM and physicochemical descriptors that define the target property manifold. QM descriptors may include, for example, the HOMO--LUMO energy gap ($E_\mathrm{gap}$), which reflects molecular electronic stability and reactivity, and the many-body dispersion energy ($E_\mathrm{MBD}$), which characterizes long-range non-covalent interactions and provides insight into molecular structure~\cite{tkatchenko12,ambrosetti2014mbd}. Physicochemical descriptors include the atom count, which controls molecular size during generation, and the synthetic accessibility (SA) score, which estimates synthetic feasibility~\cite{ertl2009sa}. A molecular discovery campaign is specified as a trajectory of navigation coordinates, $P^q=\{P_1^q, P_2^q\}$, where $q$ denotes the AL iteration and $P_1$ and $P_2$ define the selected 2D property manifold. By following this trajectory, QALPA systematically explores sparse and chemically relevant property regions while naturally supporting multi-property optimization.

Given a coordinate $P^q$, QALPA employs the E(3)-equivariant diffusion model EQDIFF to generate molecular candidates conditioned on the specified properties~\cite{edm,ho2022classifierfree,pilot}. EQDIFF is first trained on the initial dataset to learn the underlying distribution of physically realistic molecular structures by reversing a gradual noising process. During inference, the target property coordinates are supplied directly to the denoising network, allowing the model to generate three-dimensional molecular geometries that satisfy the prescribed property conditions while remaining chemically plausible. A detailed description of the diffusion architecture and training objective is provided in the following section.

Although the generated molecules are chemically valid, their atomic coordinates are not guaranteed to correspond to equilibrium structures. Consequently, each candidate can undergo geometry optimization before QM property evaluation. QALPA is compatible with any geometry optimization method; in this work we employ the ML-augmented tight-binding method EquiDTB~\cite{equidtb}, which replaces the conventional pairwise DFTB repulsive potential with a transferable many-body $\Delta_{\rm TB}$ potential parameterized through equivariant neural networks, achieving near-DFT accuracy at substantially lower computational cost. The optimized geometries provide physically meaningful structures required for reliable QM property calculations.
These geometries are subsequently evaluated by an oracle that determines whether they should be incorporated into the training dataset. To minimize computational cost, the oracle operates in two stages. First, descriptors that do not depend on geometry relaxation, such as the SA score, are evaluated to rapidly discard unsuitable candidates (only molecules with SA score lower than 5.0 are accepted). Second, the remaining molecules undergo QM property evaluation using either the all-electron electronic-structure package FHI-aims~\cite{vblum09,fhiaims2026} or EquiDTB~\cite{dftb_schnet_tcp,equidtb}, depending on the desired level of accuracy and computational efficiency. Only molecules that satisfy both the synthetic accessibility criterion and the target property tolerances are retained.

Accepted molecules are added to the existing dataset, which is subsequently used to update the diffusion model before the next AL iteration. To mitigate catastrophic forgetting during sequential learning~\cite{kirkpatrick2017}, QALPA supports either retraining from the previous checkpoint using the augmented dataset or parameter-efficient fine-tuning through low-rank adaptation (LoRA)~\cite{hu2022lora}. The active learning cycle is repeated until all navigation coordinates along the prescribed trajectory have been explored, progressively expanding the QM dataset while improving the coverage of sparsely populated regions of molecular property space.

\subsection{Equivariant diffusion model}

EQDIFF is the $E(3)$-equivariant denoising diffusion model at the core of QALPA. Its architecture is that of EQGAT-diff~\cite{eqgatdiff}, introduced for unconditional \textit{de novo} molecule generation, which QALPA adopts essentially unchanged and extends with the multi-property conditioning described below; the present work does not introduce a new diffusion network but deploys this established architecture within the active-learning loop. Each molecule is represented as an attributed graph, and atomic coordinates are generated jointly with atom types, formal charges, and bond orders. A single equivariant graph network denoises all of these modalities in parallel and, at every step, predicts the clean molecule rather than only the noise added to it. Coordinates are treated as continuous variables and follow standard Gaussian diffusion~\cite{edm}, whereas the discrete attributes (atom types, charges, bond orders) follow categorical diffusion, so the two data types are handled within one consistent framework~\cite{midi}.

\textbf{Forward noising.} During training, a forward process gradually corrupts a real molecule $x_0$ into noise over $T$ discrete timesteps. For the continuous coordinates, each step adds Gaussian noise following a variance-preserving schedule, so that the noised coordinates at time $t$ are a signal-scaled version of $x_0$ plus noise:
\begin{equation}
q(x_t \mid x_0) = \mathcal{N}\!\left(x_t;\, \bar{\alpha}_t\, x_0,\; \sigma_t^2\, \mathbf{I}\right),
\qquad
x_t = \bar{\alpha}_t\, x_0 + \sigma_t\, \varepsilon,
\qquad
\varepsilon \sim \mathcal{N}(0,\mathbf{I}),
\label{eq:forward_coords}
\end{equation}
where the signal coefficient $\bar{\alpha}_t$ decreases monotonically from $1$ (clean molecule) to $0$ (pure noise) along an adaptive cosine schedule, and the noise variance is $\sigma_t^2 = 1-\bar{\alpha}_t^2$. Each modality uses its own schedule exponent, so coordinates, bonds, and atom identities are noised at different rates; this yields a coarse-to-fine generation order in which the overall molecular shape is established first and connectivity and atom identities are resolved afterwards~\cite{eqgatdiff}. The discrete attributes are noised by a categorical process that progressively drives each class toward the empirical (marginal) class distribution $m$,
\begin{equation}
q(c_t \mid c_0) = \mathrm{Cat}\!\left(c_t;\, c_0^{\top}\, \bar{Q}_t\right),
\qquad
Q_t = \alpha_t\, \mathbf{I} + (1-\alpha_t)\, \mathbf{1}\, m^{\top},
\qquad
\bar{Q}_t = \textstyle\prod_{\tau \le t} Q_{\tau},
\label{eq:forward_cat}
\end{equation}
with transition matrices $Q_t$; the reverse step is obtained from the corresponding discrete posterior $q(c_{t-1}\mid c_t, c_0)$ (D3PM-style categorical diffusion~\cite{austin2021d3pm}), marginalized over the network's prediction of the clean attribute $c_0$.

\textbf{$E(3)$ equivariance.} Physical properties of a molecule are invariant to rigid rotations and translations, and the generative process must respect this symmetry. EQDIFF enforces it by removing the center of gravity from the input coordinates, restricting the coordinate noise $\varepsilon$ to be mean-free, and re-projecting to zero center of gravity throughout the reverse process, so that translations are factored out and the equivariant scalar and vector channels of the network make coordinate predictions rotation-equivariant~\cite{edm}. Bond predictions on the $i<j$ edges are symmetrized so that the generated adjacency is consistent~\cite{midi,eqgatdiff}.

\textbf{Reverse process and training objective.} The denoising network $f_\theta$ takes the noised coordinates, discrete attributes, and edges together with the timestep $t$ and a property context $y$ (defined below), and predicts the clean molecule for every modality simultaneously,
\begin{equation}
f_\theta\!\left(x_t, c_t, e_t, t, y\right) \;\longrightarrow\; \big(\hat{x}_0,\; \hat{p}^{\,\mathrm{atom}},\; \hat{p}^{\,\mathrm{charge}},\; \hat{p}^{\,\mathrm{bond}}\big).
\label{eq:denoiser}
\end{equation}

Training minimizes a per-modality, time-weighted reconstruction loss combining a mean-squared error on the continuous coordinates with cross-entropy terms on the categorical attributes,
\begin{equation}
\mathcal{L}
= \mathbb{E}_{t,\,q}\!\left[\, w(t)\Big(
\lambda_x\, \lVert x_0 - \hat{x}_0 \rVert^2
+ \lambda_a\, \mathrm{CE}(\hat{p}^{\,\mathrm{atom}}, a_0)
+ \lambda_b\, \mathrm{CE}(\hat{p}^{\,\mathrm{bond}}, b_0)
+ \dots \Big)\right],
\label{eq:loss}
\end{equation}
where the $\lambda$ are per-modality weights and $w(t)$ is a signal-to-noise time weighting that balances the contribution of noisy and near-clean timesteps~\cite{eqgatdiff}.

\textbf{Multi-property conditioning.} To steer generation toward a targeted region of property space, the $K$ target properties are standardized per property (using the training mean $\mu_k$ and median absolute deviation $\mathrm{mad}_k$) and concatenated into a single context vector $y \in \mathbb{R}^{K}$. This vector is embedded by a linear map $W_c$, added to the atomic scalar features $s_i$, and passed through a subsequent linear map $W'$ within the network, jointly over all $K$ properties,
\begin{equation}
y = \left[\, \frac{y^{(1)}-\mu_1}{\mathrm{mad}_1},\; \dots,\; \frac{y^{(K)}-\mu_K}{\mathrm{mad}_K} \,\right] \in \mathbb{R}^{K},
\qquad
s_i \;\leftarrow\; W'\!\left(s_i + W_c\, y\right).
\label{eq:conditioning}
\end{equation}
A single context vector therefore encodes the full multi-property target, and the network learns the joint correlations among the requested QM and physicochemical properties. Deploying the EQGAT-diff framework with this joint QM and physicochemical context is the model-level extension QALPA contributes; conditioning molecular diffusion on several quantum-mechanical properties simultaneously remains comparatively underexplored~\cite{pilot}.

\textbf{Target-directed generation.} The context is present at every training step and at every step of the reverse process, so generation uses a single conditional pass of the denoising network per step and requires no auxiliary property predictor or classifier. To sample molecules at a prescribed point in property space, the context $y$ is fixed to the standardized target values of the current trajectory point $q_i$; the reverse diffusion is then run with $y$ held at this value, so generation is steered entirely through the conditioning variable rather than an external guidance term. The framework does, however, support optional gradient-based guidance from an external energy or property model acting on the atomic coordinates, which is not used in the present workflow.

\subsection{The EquiDTB model}

In our recent work\cite{equidtb}, we demonstrated that replacing the standard pairwise DFTB repulsive potentials with an equivariant many-body $\Delta_{\rm TB}$ potential (hereafter referred to as EquiDTB25), trained using the state-of-the-art MACE architecture\cite{Batatia2022mace}, enables the prediction of multiple molecular properties with DFT-PBE0 accuracy while retaining the computational efficiency of the DFTB method.
Unlike EquiDTB25, which was trained on a subset of the QM7-X\cite{qm7x} dataset containing only molecules composed of H, C, N, and O atoms, we introduce EquiDTB26, a new model trained on the recently published QCML dataset~\cite{Ganscha2025QCML}. QCML is a large-scale quantum chemistry database designed for training and benchmarking machine learning models across diverse chemical systems comprising 79 elements. It contains approximately 33.5 million equilibrium and non-equilibrium molecular structures with up to eight non-hydrogen (heavy) atoms, for which QM properties were computed at the DFT-PBE0 level~\cite{pbe0a,adamo1999} supplemented with MBD(NL) dispersion correction\cite{mbdnl}. The dataset spans a broad range of molecular sizes, chemical compositions, charge states, and conformations, providing a substantially more diverse training domain than QM7-X.
The molecular graphs in QCML were collected from public databases and supplemented with automatically generated compounds. Initial three-dimensional geometries were optimized using the semi-empirical GFN2-xTB method\cite{bannwarth2019gfn2}, followed by conformer searches and normal-mode sampling to generate equilibrium and off-equilibrium structures. A similar workflow was employed in the construction of the QM7-X dataset. Consequently, QCML was selected to expand the chemical space covered by the original EquiDTB training set and to develop a more generalizable equivariant many-body $\Delta_{\rm TB}$ potential with a broader applicability domain than EquiDTB25.

For the present work, we constructed a subset of QCML by selecting only neutral singlet molecules ($q=0$, $M=1$) containing H, C, N, O, Na, P, S, and Cl atoms. Following the protocol used for EquiDTB25, third-order DFTB electronic energies and atomic forces were computed with the DFTB+ code\cite{hourahine2020}, and the ML corrections required to recover DFT-PBE0 accuracy were defined as,
\begin{equation}
\Delta E_{\mathrm{TB}} = E_{\mathrm{DFT}} - E_{\mathrm{DFTB3}},
\qquad
\Delta F_{\mathrm{TB}} = F_{\mathrm{DFT}} - F_{\mathrm{DFTB3}}.
\end{equation}

Compared with EquiDTB25, which was trained on approximately 500{,}000 molecular conformations from the QM7-X dataset and was limited to molecules containing up to seven non-hydrogen atoms (23 atoms in total), EquiDTB26 was developed using approximately 5.7 million molecular conformations containing up to eight non-hydrogen atoms and a maximum of 37 atoms. Of these, 4.0 million conformations were used for training set, 50{,}000 for validation set, and the remainder for the test set.
Following the original EquiDTB framework\cite{equidtb}, the EquiDTB26 model was trained using the MACE architecture to predict the energy and force corrections, $\Delta E_{\mathrm{TB}}$ and $\Delta F_{\mathrm{TB}}$, for the molecular conformations in the reduced QCML dataset. To enable a direct comparison with EquiDTB25, the same model architecture, hyperparameters, and training strategy were retained. Although an alternative training strategy was recently proposed in Ref. [\citenum{equidtbloop}], we adopted the original protocol to ensure a fair comparison between the two models.

\subsection{Benchmark molecular datasets}

\subsubsection{Small-to-large drug-like molecules}

To understand the effect of molecular dimensionality and electronic features on the performance of generative models, we considered two complementary datasets covering distinct regions of chemical space. The QM7-X\cite{qm7x} dataset comprises 41,537 small drug-like molecules containing up to seven non-hydrogen  atoms drawn from C, N, O, S, and Cl. In contrast, the Aquamarine (AQM)\cite{aqm} dataset contains 59,783 larger drug-like molecules with up to 54 heavy atoms drawn from C, N, O, F, P, S, and Cl.
In both datasets, equilibrium molecular structures were optimized using the third-order density-functional tight-binding (DFTB3) method\cite{seifert96,gaus11} combined with many-body dispersion (MBD) interactions\cite{tkatchenko12,stoehr16,mortazavi18}. However, owing to the greater conformational flexibility of larger molecules, the AQM dataset was constructed using a more extensive conformational sampling procedure. Specifically, conformers were generated using the CREST search algorithm\cite{crest2024}, which combines metadynamics simulations with genetic crossing to efficiently explore low-energy regions of the conformational landscape. Because conformations containing P atoms are relatively scarce in AQM, they were excluded from our study to avoid potential biases arising from the limited representation of P-containing molecules during model training. Consequently, the version of AQM dataset used in this work contains 58,190 conformations.

For each molecular conformation, both datasets provide approximately 40 global (molecular), local (atom-in-molecule), ground-state, and response properties obtained from QM calculations. Most of these properties were computed at the PBE0+MBD level of theory\cite{pbe0a,adamo1999,tkatchenko12,hoja2019}, employing tightly converged numeric atom-centered basis sets\cite{havu2009efficient} as implemented in the \texttt{FHI-aims} package\cite{vblum09,ren2012}. This level of theory provides an accurate description of intramolecular degrees of freedom in small organic molecules, as well as intermolecular interactions in molecular dimers, supramolecular complexes, and molecular crystals\cite{tkatchenko12,hoja2019}.
In this work, we consider only the gas-phase structures from the AQM dataset and their corresponding electronic properties. Among the available properties, we focus on two extensive quantities (many-body dispersion energy ($E_{\rm MBD}$) and molecular polarizability ($\alpha$)) and one intensive quantities (HOMO--LUMO energy gap ($E_{\rm gap}$)). Together, these properties enable us to investigate how molecular size and increasingly complex electronic environments influence the performance of property-conditioned generative models.

\subsubsection{AlloQM: Allosteric drugs}

To demonstrate the applicability of the QALPA framework to a therapeutically relevant chemical space, we assembled a QM dataset of allosteric drug molecules (Fig.~\ref{fig6}(a)). We started from the Allo-Drug subset of the AlloSteric Database 2023\cite{asd1,asd2,asd3,asd4,asd5}, which comprises 538 allosteric drugs targeting 96 allosteric proteins across 14 protein families. Among these compounds, 19 have been approved by the U.S. Food and Drug Administration, whereas the remaining molecules are distributed across different stages of the drug discovery pipeline (preclinical, Phase I, Phase II, and Phase III).
From this collection, we retrieved reliable SMILES representations for 494 of the 538 compounds. The remaining entries either lacked structural information or could not be unambiguously identified from the available metadata. After excluding non-covalent systems, three-dimensional structures were generated using RDKit, followed by conformational sampling with the CREST workflow\cite{crest2024}. This procedure yielded 477 unique molecular structures with a total of 14,423 conformers.

Because the QALPA framework integrates the EquiDTB methodology, all conformers were geometry-optimized using the EquiDTB26 model. Since this model is parameterized only for H, C, N, O, Na, P, S, and Cl atoms, the dataset was restricted to molecules containing exclusively these elements, resulting in 241 unique compounds comprising $6{,}253$ conformers.
Electronic properties were subsequently computed for every optimized conformer using DFTB3 with the \texttt{3ob} parameter set, including hydrogen corrections, together with many-body dispersion (MBD) to account for long-range van der Waals interactions. The QM property data extracted from these calculations include DFTB energy components, scalar dipole moments, HOMO--LUMO energy gaps, molecular orbital energies, and atomic Mulliken charges (Table S6 of the Supplemetary Information (SI)). The resulting dataset, denoted alloQM, comprises 6,253 conformations of allosteric drug molecules containing up to 41 non-hydrogen atoms (mean: 26) and a maximum of 78 total atoms (mean: 46.5), with elemental compositions restricted to H, C, N, O, P, S, and Cl.

\section{Results and Discussion}

\subsection{Scalability analysis of EQDIFF models}

\begin{figure}[t!]
    \centering
        \includegraphics[width=0.98\linewidth]{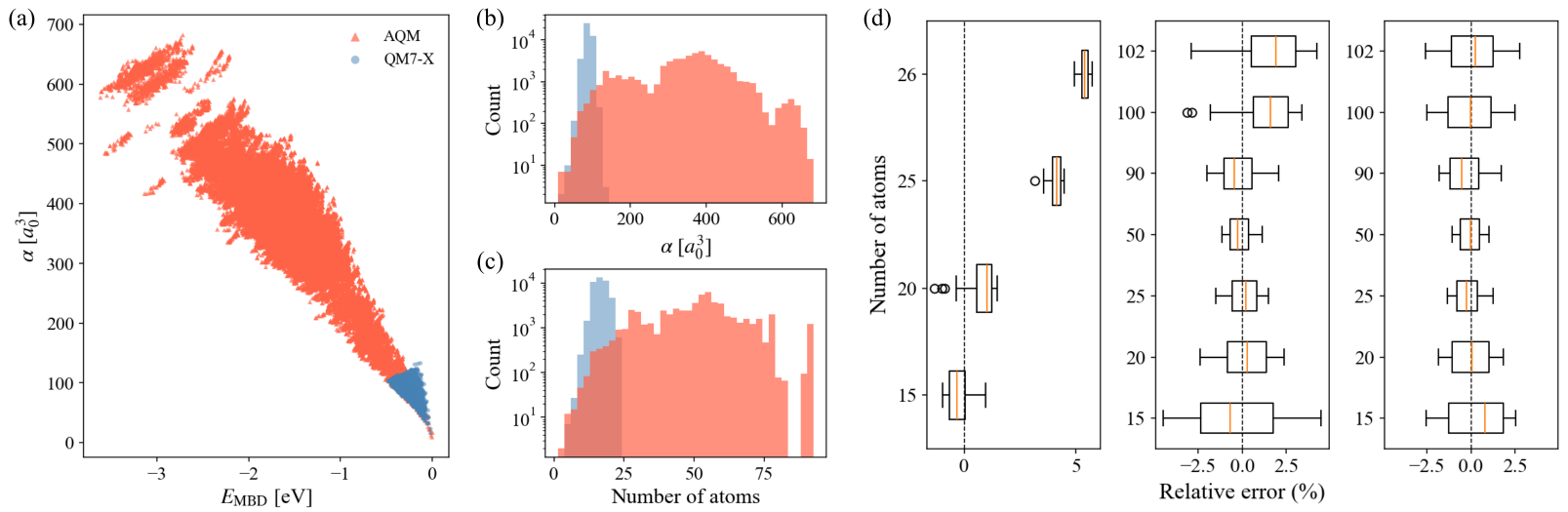}
    \caption{Scalability analysis of EQDIFF models based on extensive properties. (a) Two-dimensional property space defined by the molecular polarizability $\alpha$ and many-body dispersion energy $E_\mathrm{MBD}$ for the QM7-X (blue circles) and Aquamarine (AQM, red triangles) datasets. Both properties increase in magnitude with molecular size, with QM7-X populating the small-$\alpha$ and near-zero-$E_\mathrm{MBD}$ region, while AQM extends toward larger $\alpha$ and more negative $E_\mathrm{MBD}$. (b,c) Distributions of $\alpha$ and number of atoms, respectively, showing the smaller molecular size range covered by QM7-X and the broader range of larger molecules represented in AQM. (d) Relative error as a function of the number of atoms for molecules generated at a given target $\alpha$ using models trained on QM7-X, AQM, and the combined QM7X+AQM dataset (left to right).}
 \label{fig2}
\end{figure}

We first assess whether EQDIFF can accurately generate molecules across the range of molecular sizes encountered in drug-like chemical space. To this end, we focus on extensive QM properties whose magnitude scales with the number of atoms. Because previous studies have demonstrated that generative models can successfully design molecules with targeted molecular polarizability, $\alpha$, and many-body dispersion energy, $E_\mathrm{MBD}$,\cite{edm,medranosandonas2024inverse} we investigate the two-dimensional property space defined by these quantities, \ie the $\left(\alpha,E_\mathrm{MBD}\right)$-space. These properties provide a natural framework for evaluating the scalability of molecular generative models, as both increase systematically with molecular size.
To further investigate the effect of molecular size on model scalability, we considered the QM7-X~\cite{qm7x} and AQM~\cite{aqm} datasets, which contain small (4–23 atoms) and large (4–92 atoms) drug-like molecules, respectively. As shown in Fig.~\ref{fig2}(a--c), both $\alpha$ and $E_\mathrm{MBD}$ increase with molecular size: QM7-X molecules occupy the region with $\alpha$ values up to 120 $a_{0}^{3}$ and $E_\mathrm{MBD}$ values down to approximately $-0.5$ eV, whereas AQM spans substantially larger polarizabilities (up to 700 $a_{0}^{3}$) and more negative dispersion energies (down to $-3.6$ eV). Consequently, the two datasets cover complementary and largely non-overlapping regions of the $\left(\alpha,E_\mathrm{MBD}\right)$ space, providing a stringent benchmark for evaluating generalization across molecular sizes.

To disentangle the influence of the training distribution, we trained three EQDIFF models conditioned on $\alpha$ and $E_\mathrm{MBD}$ using QM7-X, AQM, and their combined dataset (QM7-X+AQM, 99,727 conformations), respectively. We then evaluated the relative errors of molecules generated at different target pairs of $(\alpha,E_\mathrm{MBD})$ and predefined numbers of atoms $N$ (molecular size); the corresponding values are listed in Table S1. Fig.~\ref{fig2}(d) shows the relative error in $\alpha$ as a function of the number of atoms for the 30 best-performing generated molecules at each target.
The model trained on QM7-X accurately reproduces small molecules but fails to generalize to larger systems: its relative error becomes increasingly biased and grows with the number of atoms, reflecting the absence of large-molecule statistics in the training data. In contrast, the AQM model performs best in molecular-size regions that are well represented in the training set, with the smallest errors observed for molecules containing approximately 50 atoms. The relative error increases for molecules with fewer than 20 atoms and more than 90 atoms, where the number of available training conformers is comparatively limited. Despite the scarcity of training data at the upper end of the size distribution, the AQM model successfully extrapolates into very sparse regions of the $\left(\alpha,E_\mathrm{MBD}\right)$-space, generating molecules containing up to 10 more atoms than the largest molecule present in the training set (92 atoms). This result highlights the ability of EQDIFF to extrapolate beyond the molecular size range represented during training, a capability that remains challenging for many molecular generative models.
Notably, the boxplots for the combined QM7X+AQM model remain centered around zero relative error across the entire molecular size range while exhibiting a consistently narrower spread than those of the models trained on the individual datasets. Combining the small- and large-molecule datasets therefore yields a generative model that generalizes robustly across molecular sizes, motivating its use in the property-guided molecular generation experiments presented in the following sections.

\subsection{Exploring QALPA variants for targeted molecular discovery}

\begin{figure}[t!]
    \centering
        \includegraphics[width=1.0\linewidth]{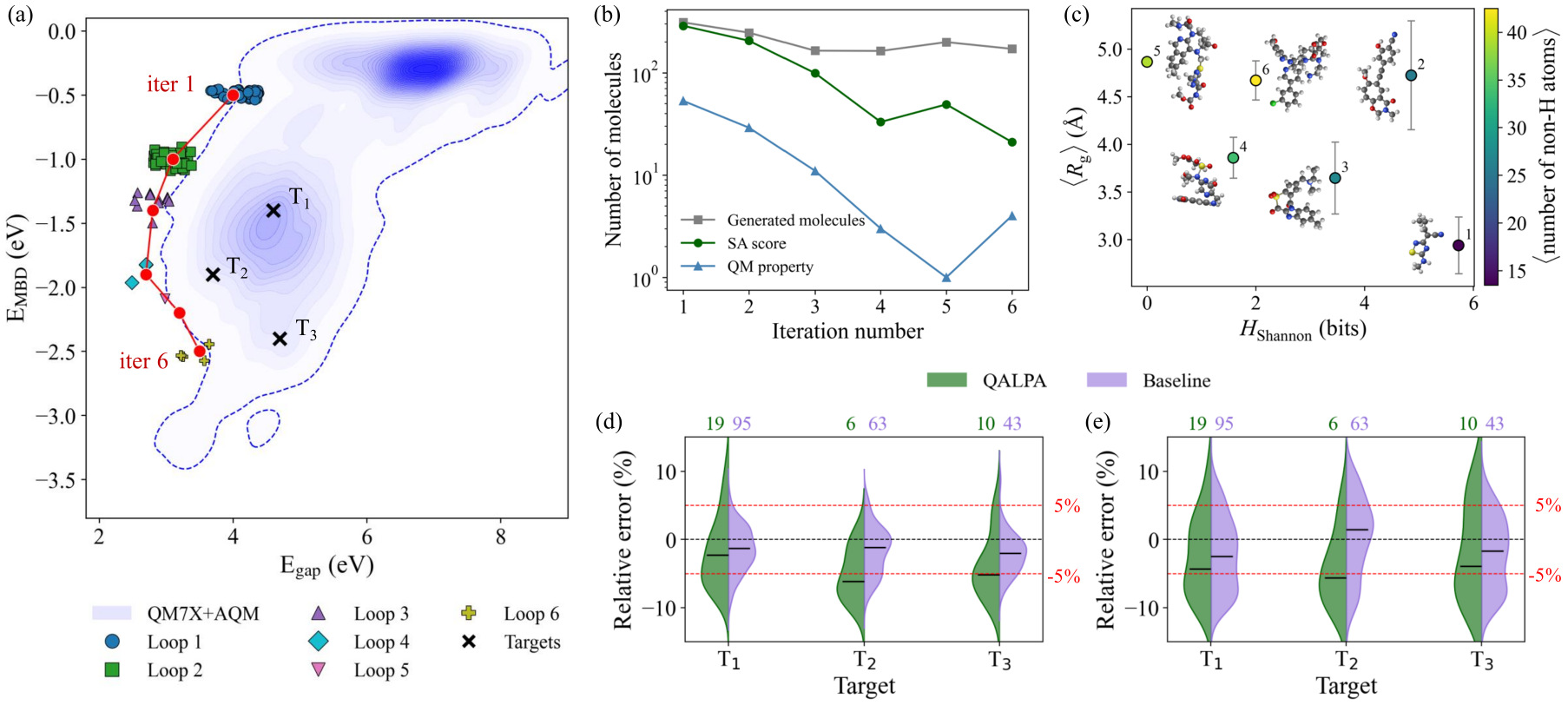}
    \caption{Property-guided generation of flexible molecules with QALPA. (a) Density plot of the property space defined by the many-body dispersion energy $E_\mathrm{MBD}$ and HOMO--LUMO energy gap $E_\mathrm{gap}$ for the combined QM7X+AQM dataset. Screened molecules generated at each navigation coordinate (red circles) are shown, with the trajectory progressing from smaller molecules with larger $E_\mathrm{gap}$ toward larger molecules with smaller $E_\mathrm{gap}$. Screened molecules at each AL iteration are represented by a distinct symbol and color. Three independent validation targets (T$_1$--T$_3$, black crosses) were defined to compare QALPA with the baseline model. (b) Number of molecules retained at each stage of the screening procedure across the AL iterations: initially generated (gray), passing the SA-score criterion (green), and passing the property-accuracy criterion (blue). (c) Average radius of gyration ($R_g$, circles) as a function of the Shannon entropy ($H$) for the screened molecular set at each AL iteration. Circle color indicates the average number of non-H atoms. Relative errors for (d) $E\mathrm{MBD}$ and (e) $E\mathrm{gap}$ for molecules screened at validation targets T$_1$--T$_3$, shown as split violin plots comparing the QALPA (green) and baseline (purple) models.}
 \label{fig3}
\end{figure}

We next evaluate the active learning (AL) strategy implemented in the QALPA framework by guiding molecular generation along a prescribed trajectory in the two-dimensional property space defined by the many-body dispersion energy, $E_\mathrm{MBD}$, and the HOMO--LUMO energy gap, $E_\mathrm{gap}$, using the combined QM7X+AQM dataset.
Fig.~\ref{fig3}(a) shows the six navigation coordinates defining the target trajectory (see red circles), located near the boundary enclosing the outermost 1\% of the data points. This trajectory spans the property landscape from small molecules with large $E_\mathrm{gap}$ and weak dispersion interactions (small $|E_\mathrm{MBD}|$) to larger molecules with smaller $E_\mathrm{gap}$ and increasingly negative $E_\mathrm{MBD}$. This trajectory simultaneously varies an intensive electronic property and an extensive long-range interaction in a very sparse region of the property space, providing a stringent test of the model ability to preserve their size-dependent correlation during molecular generation. The corresponding target property values and associated molecular sizes are listed in Table S2.
In this experiment, the QALPA workflow consists of six AL iterations, each comprising 50 training epochs, resulting in a total of 300 training epochs. At the end of every iteration, newly generated molecules that satisfy the target-property criteria (relative error < 10\% for both target properties) and present a SA score lower than 5.0 are incorporated into the training set, allowing the model to progressively refine its representation of the explored regions of chemical space. For comparison, we also trained a baseline model for the same total number of epochs (300) without active learning, \ie without augmenting the training data with newly generated molecules.

To examine how the generative model evolves throughout the AL process, we fixed the number of requested molecular graphs to 200 at each iteration. This ensures that the model attempts to generate a comparable number of candidate structures per iteration, as illustrated by the gray squares in Fig.~\ref{fig3}(b). A substantial fraction of these candidates is discarded by the SA score criterion, particularly after the third AL iteration. This trend correlates with the increasing molecular size of the target compounds, as the first three AL iterations were restricted to molecules with $N\in[23,44]$, a size range that is well represented in the combined QM7X+AQM dataset (Fig.~\ref{fig2}(c)). The subsequent property-accuracy screening  further reduces the number of candidates selected for dataset augmentation. Consequently, the final three AL iterations contribute only 3, 1, and 4 molecules, respectively.
These results demonstrate that although the generative model can produce hundreds of candidate molecules per iteration, only a small fraction satisfies both the SA score criterion and the multi-property optimization requirements needed for inclusion in the augmented dataset.

Additionally, we analyzed the radius of gyration ($R_\mathrm{g}$) and the chemical diversity of the six molecular sets generated throughout the AL process (Fig.~\ref{fig3}(c)). Chemical diversity was quantified using the Shannon entropy ($H$), computed from the distribution of atomic numbers across all molecules within each set. Higher entropy values indicate a more uniform elemental distribution and, therefore, greater chemical diversity, whereas lower values reflect compositions dominated by fewer element types. As shown in Fig.~\ref{fig3}(c), the entropy gradually decreases over successive AL iterations. This trend is explained by the progressively smaller number of molecules that pass the screening criteria, together with a reduced elemental diversity. While the first three AL iterations contain all elements represented in the combined dataset (C, N, O, F, Cl, and S), the final iterations are composed almost exclusively of C, N, O, and S atoms, with only a single molecule containing Cl in the sixth iteration.
The evolution of $R_\mathrm{g}$ further illustrates the structural diversity of the generated molecules. In particular, molecular sets 2, 5, and 6 exhibit comparable $R_\mathrm{g}$ values despite spanning different ranges of $E_\mathrm{MBD}$. At first glance, this may appear counterintuitive, as larger $E_\mathrm{MBD}$ values are often associated with larger molecules. Although this correlation generally holds because larger molecules contain more atoms, it does not directly reflect their spatial extent. The radius of gyration measures how atoms are distributed around the molecular center of mass rather than the molecular size itself. Consequently, QALPA generates both extended small molecules with relatively low $E_\mathrm{MBD}$ and compact larger molecules with higher $E_\mathrm{MBD}$. This behavior reflects the many-body nature of dispersion interactions, which depend not only on the number of atoms but also on their three-dimensional arrangement.

To quantify the evolution of model performance throughout the AL process, we defined three independent target points that were not included among the navigation coordinates (marked by the ``$\times$'' symbols in Fig.~\ref{fig3}(a)). The corresponding target property values and associated molecular sizes are listed in Table S3. Target T$_1$ (50 atoms) was selected within a densely populated region of the property manifold, whereas T$_2$ (60 atoms) and T$_3$ (75 atoms) were placed in sparsely populated regions with increasing molecular size. For each model obtained after an AL iteration, we generated 50 molecular graphs at each target and subsequently applied the screening criteria based on the SA score and property accuracy. For a fair comparison, the baseline model was evaluated using the same protocol by generating molecules every 50 training epochs.
Figs.~\ref{fig3}(d,e) present the violin plots of the prediction errors for $E_\mathrm{MBD}$ and $E_\mathrm{gap}$, respectively, considering all screened molecules accumulated over the six AL iterations. Two observations emerge. First, QALPA retains fewer molecules after screening than the baseline model (numbers above each violin plot), reflecting the increasingly demanding exploration of sparsely sampled property regions during the AL process. Second, the baseline model consistently exhibits median prediction errors closer to zero for both properties and all three targets. A likely explanation is that the baseline model is trained continuously on a fixed dataset until convergence, whereas QALPA periodically augments the training set with newly generated molecules, requiring the diffusion model to repeatedly adapt to an evolving data distribution. This continual adaptation may temporarily slow convergence, even as it progressively expands the accessible property space.
Despite this trade-off, both equivariant diffusion models generate novel molecules that simultaneously achieve high accuracy for an extensive property ($E_\mathrm{MBD}$) and an intensive property ($E_\mathrm{gap}$), with relative errors below 5\%. This result highlights the ability of the proposed framework to perform simultaneous multi-property optimization, a capability that remains challenging for many existing generative molecular design approaches.

\begin{figure}[t!]
    \centering
        \includegraphics[width=0.6\linewidth]{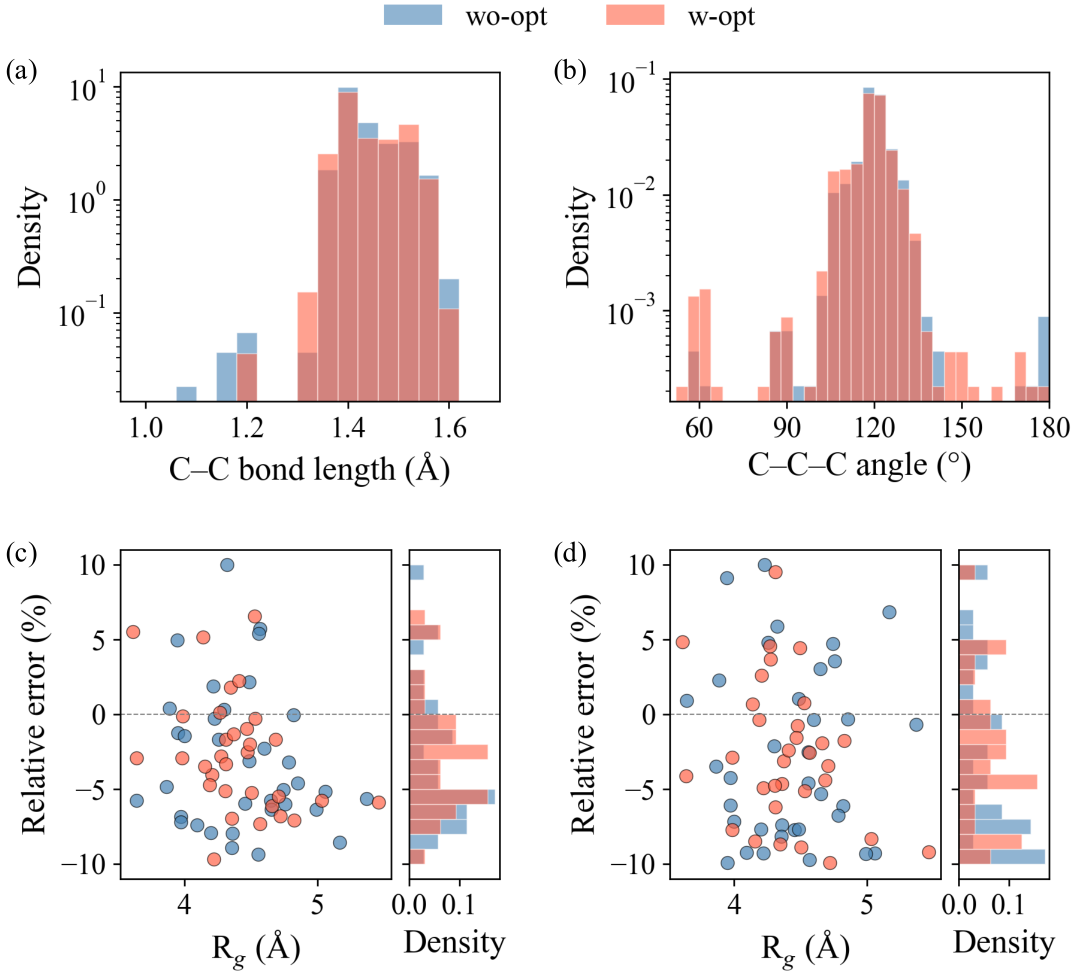}
    \caption{Influence of geometry optimization on QALPA-generated molecules. Distributions of (a) C--C bond lengths and (b) C--C--C bond angles for screened molecules generated by QALPA models trained without geometry optimization (wo-opt, blue) and with geometry optimization included (w-opt, red). Geometry optimization calculations were carried out using the DFTB3+MBD method. The distributions include all screened molecules across the six navigation coordinates. Variation of the relative error in (c) $E_\mathrm{MBD}$ and (d) $E_\mathrm{gap}$ as a function of the radius of gyration ($R_g$) for screened molecules generated at each AL iteration across the three validation targets (T$_1$--T$_3$). The lateral panels show the corresponding distributions of relative errors for the two QALPA models. }
 \label{fig4}
\end{figure}

Another important aspect that has received relatively little attention in previous generative AI studies is the geometric realism of generated molecules. Although generative models often produce chemically valid molecular graphs, their predicted three-dimensional structures may still deviate substantially from physically meaningful equilibrium geometries, potentially compromising subsequent QM property evaluation.
To investigate this effect, we trained a second QALPA model by repeating the same AL workflow described above, but considering geometry optimization of the generated molecules at each navigation coordinate before property evaluation. Geometry optimizations were performed using the DFTB3+MBD method, the same level of theory employed to generate the structures in the QM7-X and AQM datasets. For consistency with the previous analyses, we considered only the molecules that satisfied the screening criteria across the six AL iterations. The resulting molecular sets after geometry optimization are shown in Fig.~S1.

Given the chemical diversity of the combined datasets, a comprehensive analysis of all bond types and bond angles is impractical. Instead, we focus on two representative structural descriptors that are highly prevalent in drug-like molecules: C-C bond lengths and C-C-C bond angles (Figs.~\ref{fig4}(a,b)). The final datasets for this analysis comprise 101 screened molecules without geometry optimization and 99 after optimization, both spanning the same elemental composition as the initial QM7X+AQM dataset, enabling a direct comparison.
Prior to geometry relaxation, the C--C bond length distribution exhibits clear nonphysical artifacts. In particular, a population of bonds appears near 1.1~\AA{}, substantially shorter than the characteristic C$\equiv$C triple-bond length ($\sim$1.20~\AA{}), indicating unrealistic atomic arrangements. These artifacts disappear after geometry optimization, while the characteristic bond-length distributions associated with single, double, aromatic, and triple C--C bonds become clearly resolved.
Geometry optimization likewise sharpens the angle distributions, increasing the populations around chemically meaningful values such as approximately 60$^\circ$ (three-membered rings), 90$^\circ$ (four-membered rings), 109.5$^\circ$ (sp$^3$ carbon centers), and 120$^\circ$ (sp$^2$ carbon centers). Larger bond angles around 150$^\circ$ are also observed in extended molecular conformations, reflecting the flexibility of carbon backbones rather than a preferred local bonding geometry. These results indicate that a non-negligible fraction of generated molecules remains displaced from local equilibrium geometries prior to relaxation, consistent with the presence of significant residual atomic forces (Fig.~S2). 
Because QM properties are evaluated for a given molecular geometry, such structural distortions can bias the predicted properties and, consequently, the apparent accuracy of property-conditioned molecular generation. To quantify this effect, we analyzed the relative prediction errors of $E_\mathrm{MBD}$ (Fig.~\ref{fig4}(c)) and $E_\mathrm{gap}$ (Fig.~\ref{fig4}(d)) as a function of $R_\mathrm{g}$ for all screened molecules generated at the three validation targets, considering both unoptimized (35 molecules) and optimized (32 molecules) geometries. No clear correlation is observed between the prediction error and $R_\mathrm{g}$ for either property, indicating that the structural compactness of a molecule is not the primary factor controlling prediction accuracy. Nevertheless, the error distributions reveal a systematic shift toward lower relative errors for molecules generated with the QALPA model incorporating geometry optimization, indicating that structural relaxation has the potential to influence the assessment of property accuracy in generative models.
Whether such relaxation should be included in the AL loop depends on the target application. For tasks aimed at equilibrium molecular properties, geometry optimization is essential to ensure physically meaningful structures and reliable property evaluation. In contrast, for applications such as protein–ligand binding, relevant conformations may be constrained by the binding environment and do not necessarily correspond to gas-phase minima, making strict geometry relaxation less appropriate.

\subsection{Property-guided dataset augmentation of allosteric drugs}

Building on the two key components of QALPA discussed above---property-guided navigation and geometry optimization---we now demonstrate its application in a therapeutically relevant setting: the augmentation of the alloQM dataset, which contains 6,253 conformations of 241 unique allosteric drug molecules with their corresponding 16 local and global QM properties (see Methods).
As a proof-of-concept, we integrate the EquiDTB framework into QALPA to obtain optimized molecular structures at a level of accuracy comparable to PBE0+MBD (quality reference geometries). Electronic properties are consistently evaluated at the DFTB3+MBD level using the 3ob Slater-Koster parameter set, which has been shown to perform reliably for a wide range of organic molecular systems. These calculations are performed using the established EquiDTB workflow\cite{equidtb}, enabling a fully consistent and computationally efficient pipeline for structure refinement and property evaluation within the QALPA framework.

\begin{figure}[t!]
    \centering
        \includegraphics[width=0.5\linewidth]{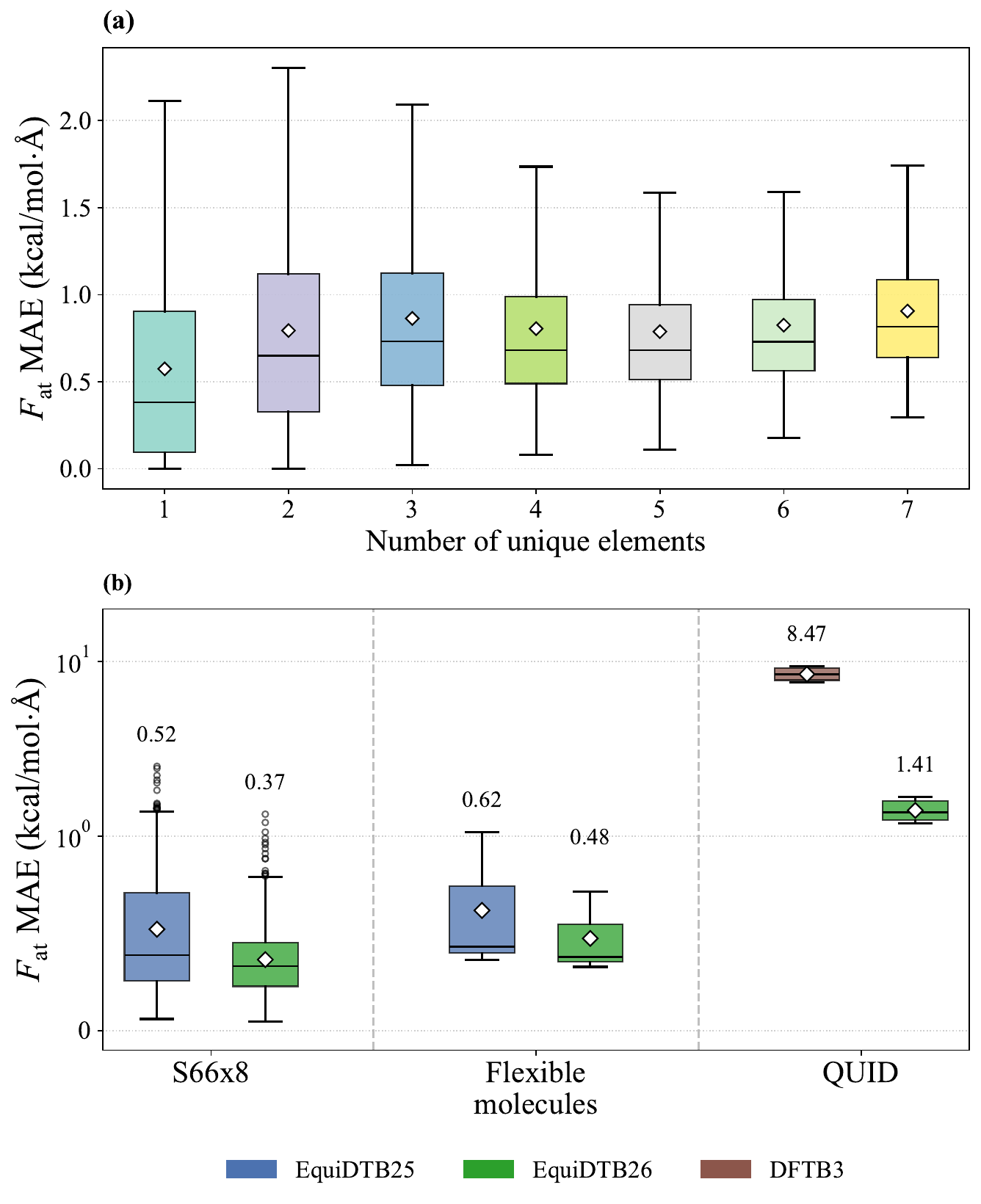}
    \caption{Benchmarking the EquiDTB26 model. (a) Boxplots of the mean absolute errors (MAEs) for atomic forces ($F_{\rm at}$) predicted by the best-performing EquiDTB26 model across molecules of increasing chemical complexity, ranging from molecules containing a single unique element to those containing seven unique elements. The analysis uses a subset of the QCML dataset comprising $\approx$5.7 million equilibrium and non-equilibrium conformations of small molecules containing up to eight non-hydrogen atoms. (b) Comparison of EquiDTB26, the previously developed EquiDTB25 model, and standard DFTB3 for force predictions across three validation benchmarks involving molecular dimers (S66x8 and QUID) and flexible molecules (paracetamol, zaprinast, and ligand 2Q5k). Boxplots show the force MAEs on a logarithmic scale, with the corresponding mean values indicated above each box.}
\label{fig5}
\end{figure}

\subsubsection{Validating the EquiDTB26 model}

This work introduces EquiDTB26, an updated equivariant many-body $\Delta_{\rm TB}$ potential trained on the substantially larger and chemically more diverse QCML dataset (Fig. S4)~\cite{Ganscha2025QCML}. Five models were trained using progressively larger subsets of QCML, ranging from 0.5 to 4.0 million (M) equilibrium and non-equilibrium conformations of small organic molecules containing up to eight non-hydrogen atoms composed of C, N, O, Na, P, S, and Cl.
We first evaluated the ability of these models to learn the energy ($\Delta E_{\mathrm{TB}}$) and atomic force ($\Delta F_{\mathrm{TB}}$) corrections required to reproduce DFT-PBE0 reference calculations. Among the five models, the 4M model (hereafter referred to as EquiDTB26) consistently achieved the highest accuracy on both the QCML test set (Table S4) and all additional validation benchmarks (Table S5). EquiDTB26 yields mean absolute errors (MAEs) of 0.058 and 0.059 kcal/mol/atom for $\Delta E_{\mathrm{TB}}$ on the training ($\sim$4.0 M conformations) and test ($\sim$1.7 M conformations) sets, respectively. The corresponding MAEs for $\Delta F_{\mathrm{TB}}$ are 0.891 and 0.905 kcal/mol/\AA{}. The close agreement between training and test errors demonstrates excellent generalization to previously unseen molecular conformations, with no evidence of overfitting.
Although these errors are approximately twice those of EquiDTB25, the comparison should be interpreted with caution because that model was trained on the chemically less diverse QM7-X dataset, which contains only molecules composed of H, C, N, and O atoms. In contrast, QCML spans a much broader chemical space, including additional elements, more diverse bonding environments, and significantly greater structural complexity. Despite this substantially more challenging learning task, EquiDTB26 maintains low prediction errors across the expanded chemical domain (Fig.~\ref{fig5}(a)), demonstrating the robustness and transferability of the many-body $\Delta_{\rm TB}$ potential.

We next evaluate the ability of EquiDTB26 to predict atomic forces, $F_\mathrm{at}$, for systems that were not represented during training, including molecular dimers of different sizes and non-equilibrium conformations of flexible molecules (Fig.~\ref{fig5}(b)). Because long-range dispersion interactions play a central role in these systems, all calculations include the many-body dispersion (MBD) correction.
Note that all reference data used for the validation benchmarks were also computed at the PBE0+MBD level of theory.
We first consider the S66$\times$8 benchmark dataset\cite{rezac2011s66x8}, which comprises equilibrium and non-equilibrium configurations of small non-covalent molecular dimers. Relative to EquiDTB25, EquiDTB26 reduces the force MAE from 0.52 to 0.37 kcal/mol/\AA{}, indicating that the broader and more diverse training set substantially improves the description of intermolecular interactions of S66$\times$8 dimers.
To assess transferability to larger non-covalent complexes, we further evaluated EquiDTB26 on the recently introduced QUID dataset\cite{puleva2025quid}, which contains chemically diverse large molecular dimers. Because EquiDTB26 was trained only on a subset of the chemical space represented in QUID, atomic forces were computed for the 32 equilibrium and non-equilibrium dimers that fall within its chemical domain (\ie F2B1, F2B2, F2I1, and F2I2 dimers). For these systems, EquiDTB26 achieves a force MAE of 1.39 kcal/mol/\AA{}, compared with 8.46 kcal/mol/\AA{} for the conventional DFTB3 method employing a pairwise repulsive potential. This approximately sixfold reduction in prediction error highlights the substantial improvement provided by our ML correction. Moreover, the improved force accuracy is consistently maintained across the full range of intermolecular separation factors, $q$ (Fig. S5).
Finally, we evaluated EquiDTB26 on thermal conformations of three flexible molecules (paracetamol, zaprinast, and ligand 2Q5k) previously used to benchmark EquiDTB25 against other tight-binding methods\cite{equidtb}, including DFTB3 and GFN2-xTB. Overall, EquiDTB26 further improves upon EquiDTB25, reducing both the spread of the force-error distributions  and the averaged MAE from 0.62 to 0.48 kcal/mol/\AA{} (Fig.~\ref{fig5}(b)). The largest improvements are observed for zaprinast and ligand 2Q5k, whose MAEs decrease from 1.06 to 0.72 kcal/mol/\AA{} and from 0.43 to 0.33 kcal/mol/\AA{}, respectively. Whereas the error for paracetamol remains essentially unchanged, increasing only slightly from 0.37 to 0.38 kcal/mol/\AA{}. Correlation plots of the predicted atomic forces for each molecule further confirm the improved performance of EquiDTB26 (Fig. S6).
Taken together, these results demonstrate that expanding the training data with the more chemically diverse QCML dataset substantially enhances the transferability and scalability of the equivariant many-body $\Delta_\mathrm{TB}$ potential. EquiDTB26 therefore provides an accurate and computationally efficient model for geometry optimization and property evaluation across a significantly broader region of chemical space than its predecessor.

\subsubsection{Allosteric-drug dataset augmentation}

\begin{figure}[t!]
    \centering
        \includegraphics[width=1.0\linewidth]{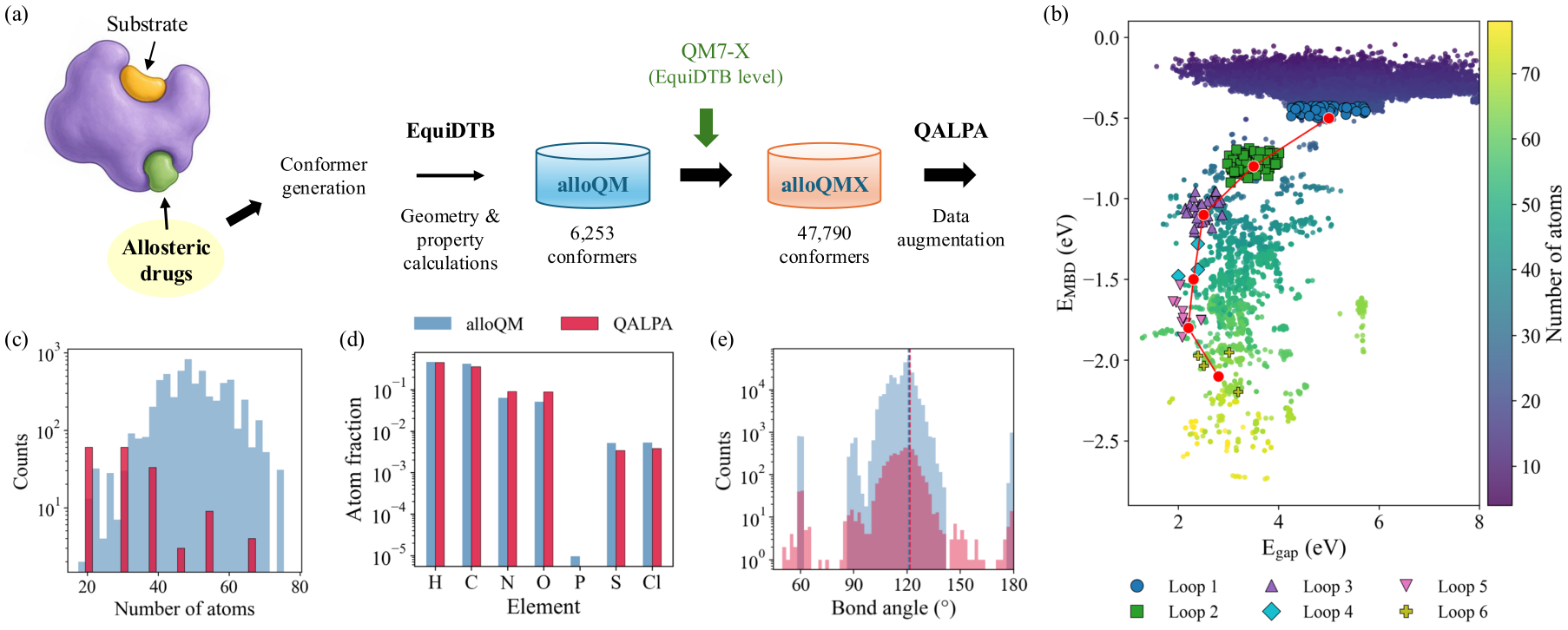}
    \caption{Application of the QALPA framework to augment a quantum-mechanical (QM) dataset of allosteric drug molecules. (a) Workflow: conformers of allosteric drugs are generated and processed using EquiDTB for geometry optimization and property evaluation to construct the alloQM dataset ($6{,}253$ conformers), which is combined with QM7-X to form the alloQMX training set ($47{,}790$ conformers). (b) Explored property space defined by the many-body dispersion energy $E_{\mathrm{MBD}}$ and HOMO--LUMO energy gap $E_{\mathrm{gap}}$, with points colored according to the number of atoms. Red circles indicate the navigation coordinates along the planned trajectory (AL iterations 1--6). The chemical and structural diversity of the screened molecules generated across the six iterations is compared with that of alloQM in terms of (c) number-of-atom distributions, (d) elemental composition, and (e) all angle distributions. The dashed line at 120.7$^\circ$ indicates the most frequently occurring angle in the alloQM dataset.}
\label{fig6}
\end{figure}

We finally demonstrate the ability of the QALPA framework to explore sparsely sampled regions of QM property space by augmenting a dataset of allosteric drug molecules (Fig.~\ref{fig6}). Starting from conformers of allosteric drugs processed using the EquiDTB workflow with the EquiDTB26 model, we constructed the alloQM dataset, initially comprising 6,253 conformers (see Methods).
To improve the robustness of the generative model, the alloQM dataset was combined with QM7-X, yielding the alloQMX dataset containing 47,790 conformers composed of H, C, N, O, P, S, and Cl atoms. As demonstrated previously for the AQM dataset, enriching the training data with additional molecular fragments and local chemical environments improves the performance of the AL workflow by broadening the underlying chemical distribution.
To ensure a consistent computational reference, all QM7-X geometries were first optimized using the EquiDTB workflow, after which their QM properties were recomputed at the same DFTB3+MBD level employed for alloQM (Fig.~\ref{fig6}(a)). We selected $E_{\mathrm{MBD}}$ and $E_{\mathrm{gap}}$ as the navigation coordinates and defined a six-step trajectory through this 2D property space (Fig.~\ref{fig6}(b)). The target property pairs associated with each AL iteration are listed in Table S7.

Throughout the AL iterations, QALPA efficiently generates molecules in densely populated regions of the property space that pass the screening criteria: SA score < 5.0 and relative error < 15\% for both target properties. For example, the first three navigation coordinates yield a total of 153 conformers containing 20, 32, and 39 atoms, respectively (Fig.~\ref{fig6}(c)). As the trajectory progresses toward increasingly sparse regions corresponding to larger molecules, molecular generation becomes more challenging, and only 16 conformers containing 44, 56, and 68 atoms are produced across the final three navigation coordinates. Nevertheless, these newly generated structures effectively enrich regions of the property space that are poorly represented in the original alloQM dataset, demonstrating the ability of QALPA to perform targeted data augmentation where additional samples are most needed.

Importantly, the generated molecules preserve the overall chemical characteristics of the original dataset while extending its coverage. The elemental composition closely matches that of alloQM across H, C, N, O, S, and Cl atoms (Fig.~\ref{fig6}(d)), whereas phosphorus, which is only sparsely represented in the initial dataset, is correspondingly rare among the generated molecules.
Likewise, the angle distribution exhibits a pronounced maximum near $121.4^\circ$, closely matching the reference distribution of alloQM ($120.7^\circ$) and indicating that the generated structures remain geometrically realistic (Fig.~\ref{fig6}(e)). The Shannon entropy also increased from 1.11 to 1.20, suggesting an increase in the chemical diversity of the new molecular set. Moreover, new angle configurations emerge in the generated molecules, consistent with the extension of the dataset into previously unsampled regions of conformational space.
These results demonstrate that property-guided molecular generation, combined with EquiDTB-based geometry optimization, produces chemically plausible structures that successfully expand the alloQM dataset along a prescribed trajectory in QM property space.

\section{Conclusions}

In this work, we introduced QALPA, a property-guided diffusion framework that explores chemical space by generating molecules conditioned on target QM property manifolds. QALPA combines generative diffusion models with an active learning workflow in which newly generated molecules are iteratively refined through geometry optimization, QM property evaluation, and dataset augmentation, enabling continuous improvement of the underlying generative model.
Our results demonstrate that integrating datasets spanning both small molecules (QM7-X) and larger drug-like compounds (AQM) substantially improves the transferability of the model across molecular sizes, overcoming the limited extrapolation observed when training exclusively on small molecules. Under multi-property guidance in the $\left(E_{\mathrm{MBD}}, E_{\mathrm{gap}}\right)$-space, QALPA consistently generated synthetically accessible and chemically diverse molecules with increased sizes and high property accuracy in sparse regions. 
Furthermore, incorporating DFTB-based geometry optimization effectively corrected the generated molecular structures, yielding chemically realistic bond lengths and bond angles that are essential for reliable QM property evaluation.
At the same time, geometry optimization introduced a measurable trade-off between structural realism and agreement with the target property manifold, highlighting that the optimal level of structural refinement depends on the specific multi-property optimization problem being addressed.

As a proof of concept, we further demonstrated the capability of QALPA by augmenting alloQM, a dataset introduced in this work that comprises QM properties for $6{,}253$ equilibrium conformers of 241 unique allosteric drug molecules. By coupling QALPA with the EquiDTB method, we successfully navigated the $\left(E_{\mathrm{MBD}}, E_{\mathrm{gap}}\right)$-space and generated 169 additional molecules spanning a diverse range of molecular sizes and elemental compositions. The resulting structures exhibit chemically realistic geometries while retaining DFTB3+MBD accuracy in the predicted electronic properties, illustrating the ability of the framework to efficiently enrich sparse regions of molecular property space.
More broadly, these results show that integrating generative AI with efficient ML/QM methods provides a practical strategy for augmenting sparse QM datasets and exploring complex property landscapes.
In this context, QALPA serves as a general framework for accelerating the exploration of chemical space by generating focused candidate sets at prescribed property coordinates for downstream structural refinement and expert-guided molecular design. Owing to its modular architecture, the framework can be readily combined with next-generation AI-based electronic-structure methods, providing a pathway toward improved efficiency, accuracy, scalability, and transferability while enabling a more sustainable exploration of previously inaccessible regions of chemical space.

\section*{Acknowledgments}

LMS thanks Prof. Gianaurelio Cuniberti and the members of the Chair of Materials Science and Nanotechnology at TUD Dresden University of Technology for the support during the development of this work.
We thank the Center for Information Services and High-Performance Computing (ZIH) at TU Dresden for providing the computational resources and technical support.

\section*{Author Contributions}
The work was initially conceived by LMS and JC and designed with contributions from MH and ZE.
MH developed the active learning workflow and the QALPA GitHub repository, with assistance from LMS.
JC contributed to the implementation of EQDIFF within the QALPA framework.
LMS performed model training across multiple applications and analyzed the corresponding performance.
ZE generated the dataset used to develop the EquiDTB26 model and subsequently evaluated its performance.
LMS supervised the project and reviewed all stages of the work.
LMS drafted the original manuscript and designed the figures, with input from MH, JC, and ZE.
All authors discussed the results and contributed to the final version of the manuscript.

\section*{Data Availability}

The scripts required to run the QALPA framework are available in the QALPA GitHub repository (https://github.com/lmedranos/qalpa).
The trained generative models and additional benchmarking datasets used in this study are also provided in the QALPA repository.
The dataset files containing the target energies $\Delta E_{\rm TB}$ and atomic forces $\Delta \mathbf{F}_{\rm TB}$ for the QCML subset are available in the EquiDTB GitHub repository (https://github.com/lmedranos/EquiDTB).

\section*{Conflicts of interest}
There are no conflicts to declare.

\bibliography{eqmgen}

\end{document}


\fancyhead{}
 \renewcommand{\headrulewidth}{1pt}
 \renewcommand{\footrulewidth}{1pt}
 \setlength{\arrayrulewidth}{1pt}
 \setlength{\columnsep}{6.5mm}
 \renewcommand{\figurename}{Fig.~S\!\!}
 \renewcommand{\tablename}{Table~S\!\!}

 \begin{center}
 \noindent\LARGE{Supplementary Information (SI) for:}
 \vspace{0.3cm}

 \noindent\LARGE{\textbf{QALPA: Property-guided diffusion modeling for efficient exploration of chemical spaces of flexible molecules}}
 \vspace{0.6cm}

 \noindent\large{\textbf{Michael Hanna\textit{$^{1}$}, Julian Cremer\textit{$^{2}$}, Zekiye Erarslan\textit{$^{1,3,4}$}, and Leonardo Medrano Sandonas,$^{\ast}$\textit{$^{3,4,5}$}}} \vspace{0.5cm}

 \noindent{\textit{$^{1}$~Faculty of Computer Science, TUD Dresden University of Technology, 01062 Dresden, Germany}}\\[0.6em]
  \noindent{\textit{$^{2}$~Machine Learning \& Computational Sciences, Pfizer Worldwide R\&D, Berlin, Germany }}\\[0.6em]
   \noindent{\textit{$^{3}$~Center for Advanced Systems Understanding (CASUS), Conrad-Schiedt-Straße 20, Görlitz 02826, Germany }}\\[0.6em]
    \noindent{\textit{$^{4}$~Helmholtz Zentrum Dresden-Rossendorf, Bautzner Landstraße 400, Dresden 01328, Germany}}\\[0.6em]
  \noindent\small{\textit{$^{5}$~Institute for Materials Science and Max Bergmann Center of Biomaterials, TUD Dresden University of Technology, 01062 Dresden, Germany}}\\[1em]
  \noindent{$^{\ast}$~Corresponding author: Leonardo Medrano Sandonas (\texttt{l.medrano-sandonas@hzdr.de})}
 \end{center}

\vspace{0.5in}
\tableofcontents

 
\clearpage


\section{Scalability test}

\begin{table*}[h]
  \centering
    \caption{List of validation targets in the property space defined by the molecular polarizability $\alpha$ and many-body dispersion energy $E_{\rm MBD}$ for EQDIFF models trained on QM7-X, AQM, and the combined QM7-X+AQM datasets. 
  }

    \begin{tabular}{lllllllll}
    \hline\hline
  Dataset & Property  & 1 & 2 & 3 & 4 & 5 & 6 & 7  \\
\hline
 \multirow{3}{2.5cm}{QM7-X} & $N$ & 15 & 20 & 25 & 26 & -- &--  & --  \\
    &  $\alpha$  & 83.5278 & 97.0245 & 110.5213 & 113.2206 & -- & -- & --  \\
   & $E_{\rm MBD}$  & -0.2332 & -0.3485	 & -0.4639	 & -0.4869	& -- & -- & -- \\
 \hline
 \multirow{3}{2.5cm}{AQM} & $N$ & 15 & 20 & 25 & 50 & 90 & 100 & 102  \\
    &  $\alpha$  &  102.2349 & 136.4424 & 170.6499 & 341.6874 & 615.3474 & 683.7624 &  697.4454 \\
   & $E_{\rm MBD}$  & -0.1761	  & -0.3596	  & -0.5431	& -1.4605 & -2.9283 & -3.2952	 & -3.3686	  \\
 \hline
 \multirow{3}{2.5cm}{QM7X+AQM} & $N$ & 15 & 20 & 25 & 50 & 90 & 100 & 102  \\
    &  $\alpha$  &  80.7339 & 117.3736 & 154.0133 & 337.2118 & 630.3294 & 703.6088 &  718.2647 \\
   & $E_{\rm MBD}$  & -0.2043  & -0.3844  & -0.5644 & -1.4645 & -2.9048 & -3.2648  & -3.3368  \\
    \hline\hline 
  \end{tabular}
  \label{table1}
\end{table*}

\clearpage 

\section{Additional QALPA results on combined dataset}

\begin{table*}[h]
  \centering
    \caption{List of navigation coordinates in the property space defined by the many-body dispersion energy $E_{\rm MBD}$ and HOMO--LUMO energy gap $E_{\rm gap}$ for QALPA models trained on the combined QM7-X+AQM dataset, considering either with or without geometry optimization at each active learning iteration.
  }

    \begin{tabular}{lllllll}
    \hline\hline
   Property  & 1 & 2 & 3 & 4 & 5 & 6   \\
   \hline
  $N$ & 23 & 41 & 44 & 57 & 69 & 78   \\
      $E_{\rm MBD}$  & -0.5 & -1.0 & -1.4 & -1.9 & -2.2 & -2.5  \\
    $E_{\rm gap}$  & 4.0 & 3.1 & 2.8 & 2.7 & 3.2 & 3.5  \\

    \hline\hline 
  \end{tabular}
  \label{table2}
\end{table*}

\begin{table*}[h]
  \centering
    \caption{List of additional validation targets (T) in the property space defined by the many-body dispersion energy $E_{\rm MBD}$ and HOMO--LUMO energy gap $E_{\rm gap}$ for QALPA models trained on the combined QM7-X+AQM dataset, considering either with or without geometry optimization at each active learning iteration. 
  }

    \begin{tabular}{lllllll}
    \hline\hline
   Property  & T$_1$ & T$_2$ & T$_3$   \\
   \hline
  $N$ & 50 & 60 & 75    \\
      $E_{\rm MBD}$  & -1.4& -1.9 & -2.4   \\
    $E_{\rm gap}$  & 4.6 & 3.7 & 4.7   \\

    \hline\hline 
  \end{tabular}
  \label{table3}
\end{table*}

\begin{figure}[h]
    \centering
    \includegraphics[width=0.9\linewidth]{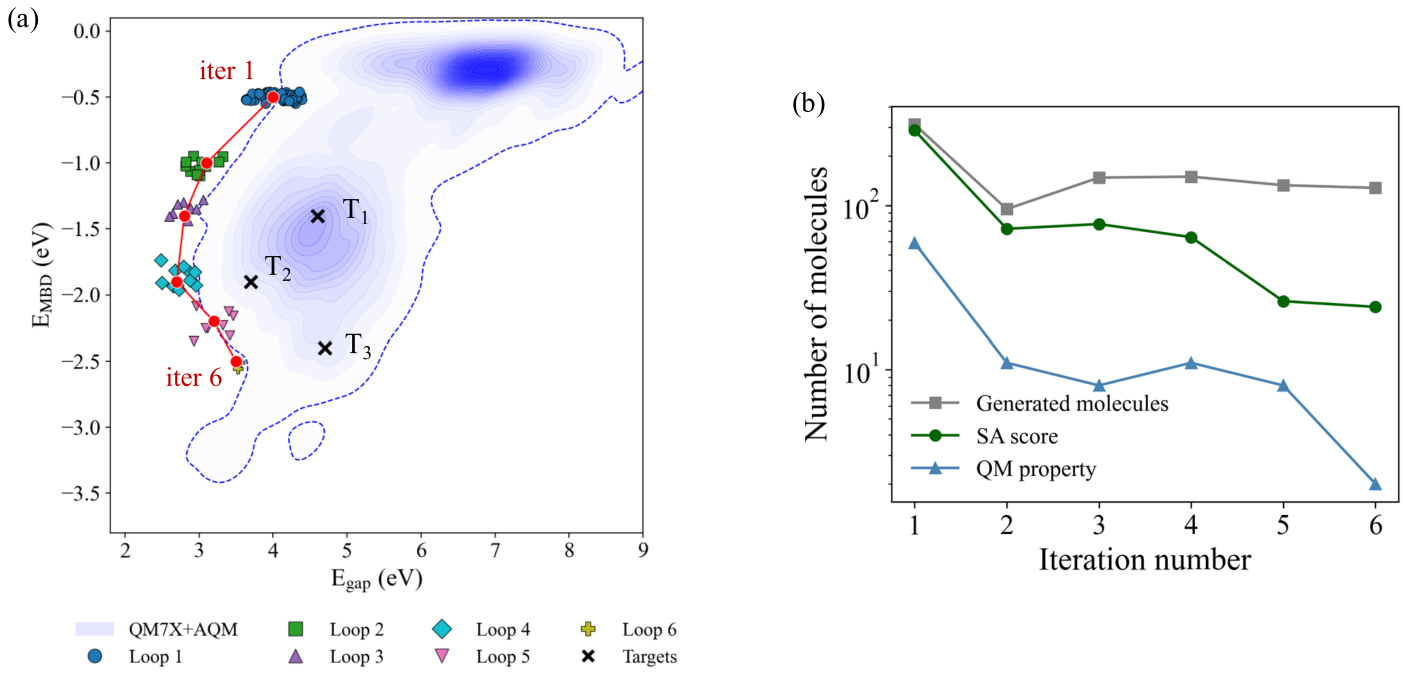}
    \caption{Property-guided generation of flexible molecules with QALPA, considering geometry optimization at every AL loop. (a) Density plot of the property space defined by the many-body dispersion energy $E_\mathrm{MBD}$ and HOMO--LUMO energy gap $E_\mathrm{gap}$ for the combined QM7X+AQM dataset. Screened molecules generated at each navigation coordinate (red circles) are shown, with the trajectory progressing from smaller molecules with larger $E_\mathrm{gap}$ toward larger molecules with smaller $E_\mathrm{gap}$. Screened molecules at each AL iteration are represented by a distinct symbol and color. Three independent validation targets (T$_1$--T$_3$, black crosses) were defined to compare QALPA model with the baseline model. (b) Number of molecules retained at each stage of the screening procedure across the AL iterations: initially generated (gray), passing the SA-score criterion (green), and passing the property-accuracy criterion (blue).}
    \label{sfig1}
\end{figure}

\begin{figure}[h]
    \centering
    \includegraphics[width=0.8\linewidth]{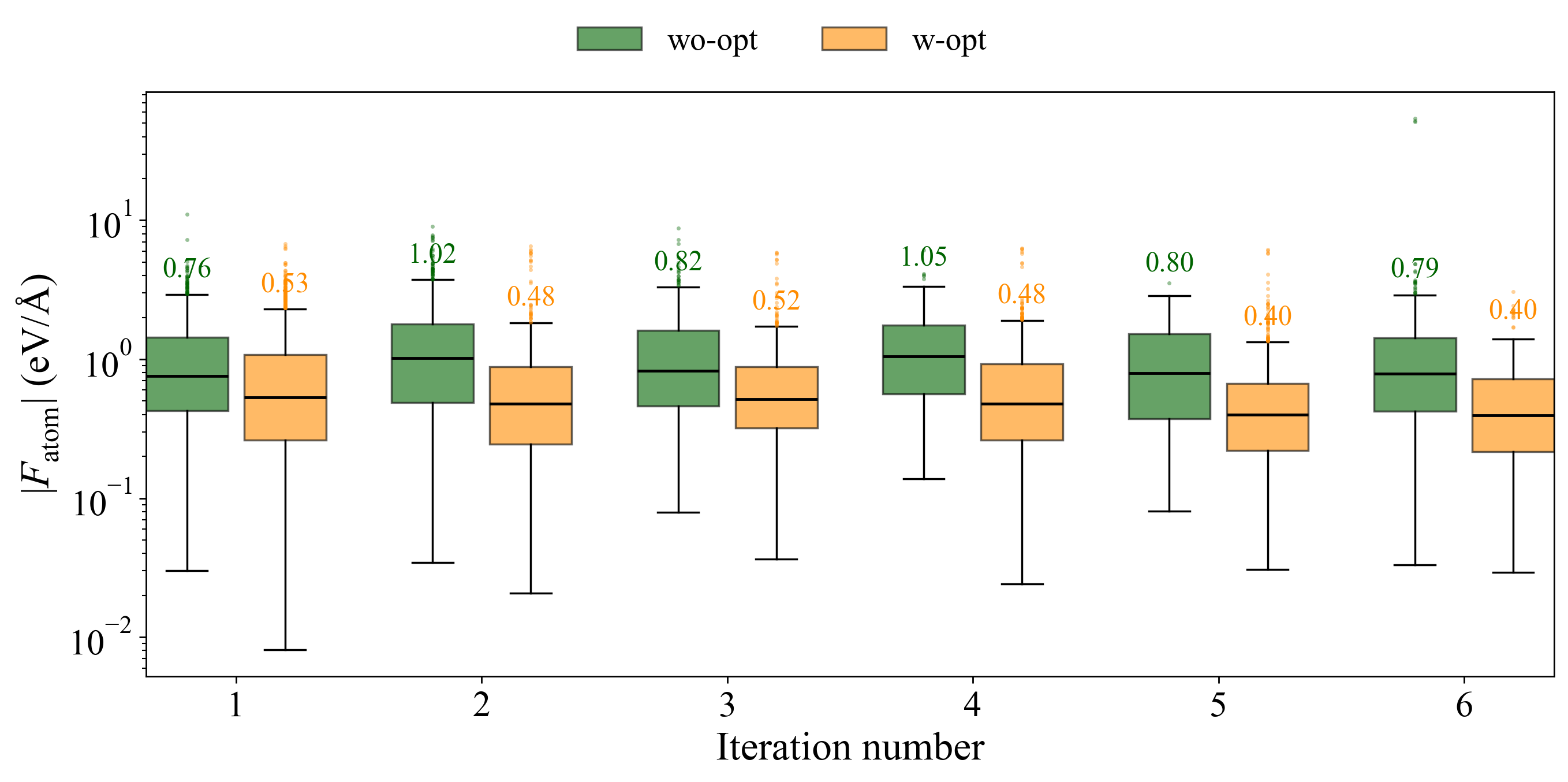}
    \caption{Comparison of the atomic-force ($F_{\rm atom}$) distributions at each iteration of the QALPA active learning workflow, with and without geometry optimization. Mean values are reported above each boxplot.}
    \label{sfig2}
\end{figure}

\begin{figure}[h]
    \centering
    \includegraphics[width=0.8\linewidth]{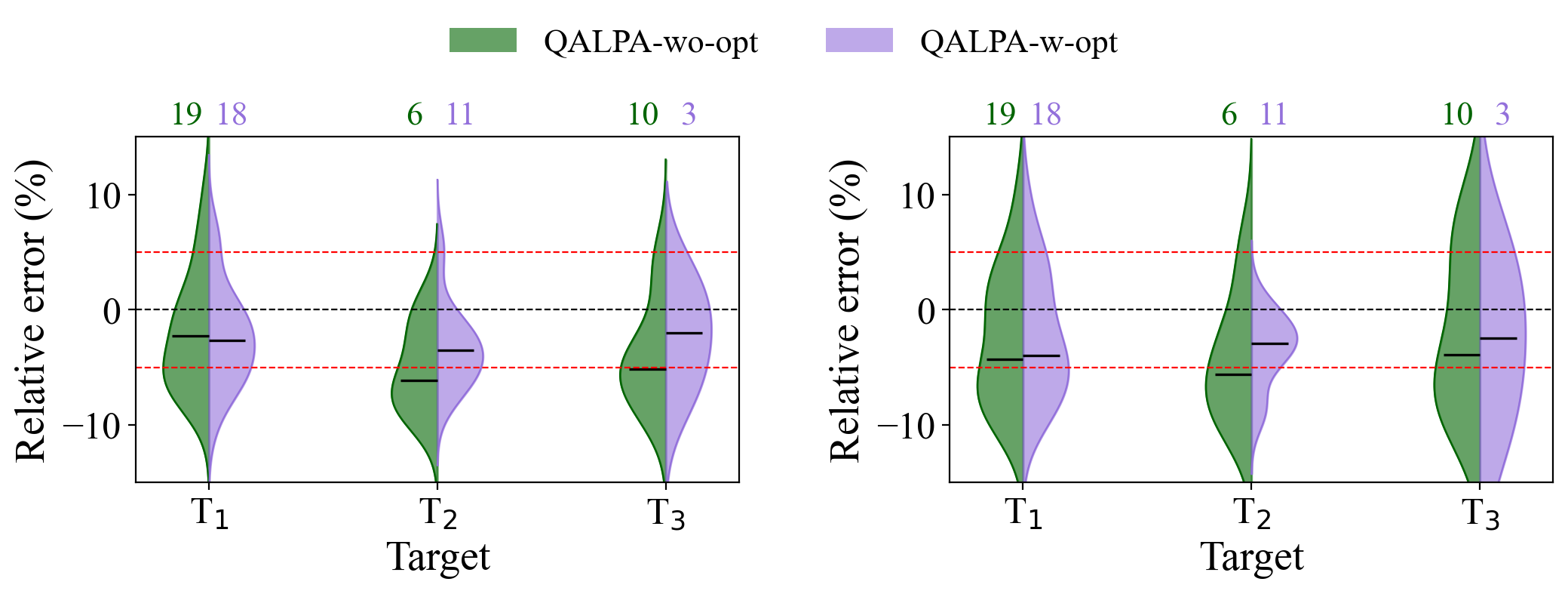}
    \caption{Property-guided generation of flexible molecules with QALPA, considering geometry optimization at every AL loop. Relative errors for (left panel) $E\mathrm{MBD}$ and (right panel) $E\mathrm{gap}$ for molecules screened at validation targets T$_1$--T$_3$, shown as split violin plots comparing the QALPA (green) and baseline (purple) models.}
    \label{sfig3}
\end{figure}

\clearpage

\section{Validating EquiDTB26 model}


\begin{figure}[b!]
    \centering
    \includegraphics[width=0.9\linewidth]{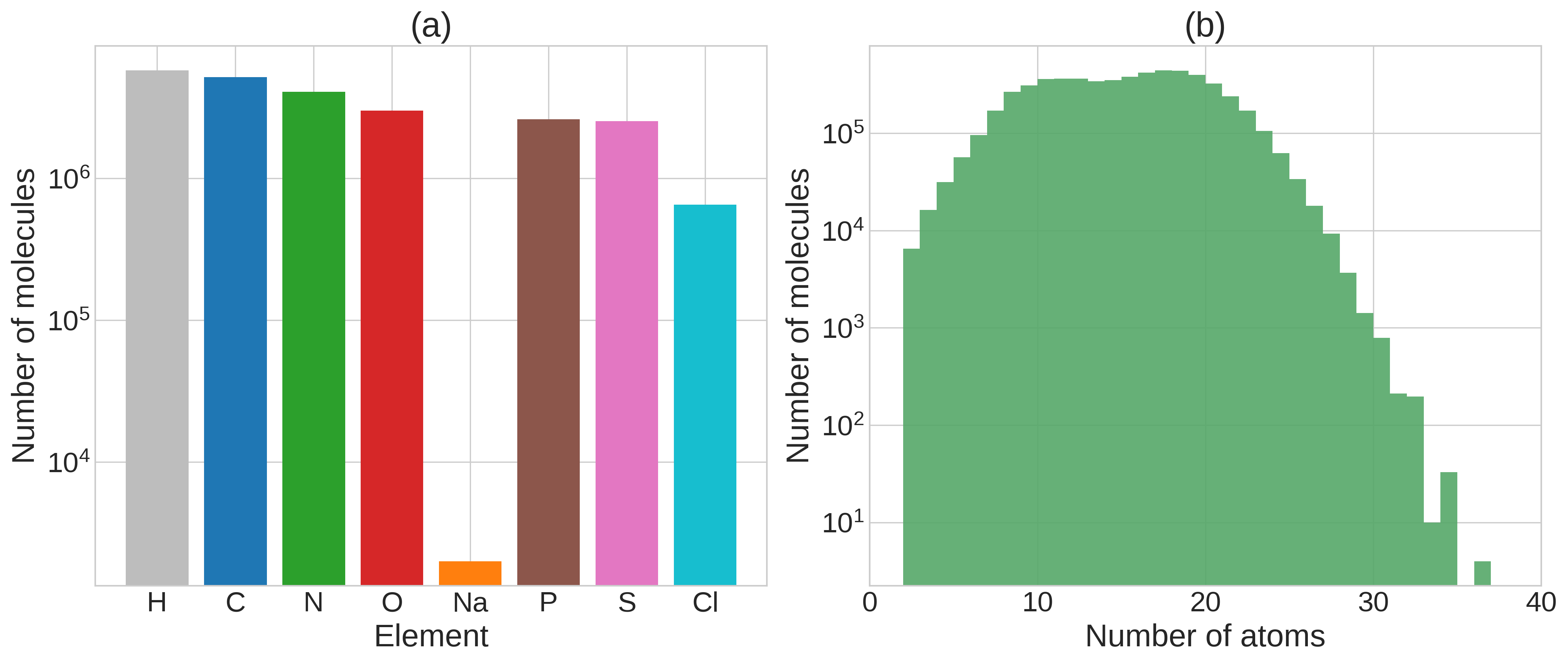}
    \caption{Elemental composition and molecular-size distribution of the QCML dataset. (a) Number of molecular conformations containing each chemical element. (b) Distribution of the number of atoms per molecular conformation.}
    \label{fig:S4}
\end{figure}

Fig.~S\ref{fig:S4} illustrates the chemical diversity of the QCML training dataset. The dataset spans molecules containing 2 to 37 atoms, with the majority containing between 8 and 20 atoms. This broader elemental coverage and wider molecular-size distribution are expected to improve the transferability of the learned force correction to larger and chemically more diverse systems.

To investigate the effect of training-set size, we trained a series of equivariant many-body $\Delta_{\rm TB}$ potentials using progressively larger subsets of QCML containing 500k, 1M, 2M, 3M, and 4M molecular conformations. All models were evaluated on the QCML test set using their predicted energies and atomic forces.
As summarized in Table~S\ref{tab:qcml_model_mae}, increasing the number of QCML training conformations leads to an overall reduction in both energy and force errors. The test-set energy MAE decreases from 0.132~kcal/mol/atom for the model trained on 500k conformations to 0.059~kcal/mol/atom for the model trained on 4M conformations. Over the same range, the force MAE decreases from 1.716 to 0.905~kcal/mol/\AA{}. These results demonstrate that increasing the QCML training set improves the accuracy of the learned energy and force corrections within the QCML chemical space. Hereafter, we refer to the best-performing model as EquiDTB26.

\begin{figure}[t!]
    \centering
    \includegraphics[width=1.0\linewidth]{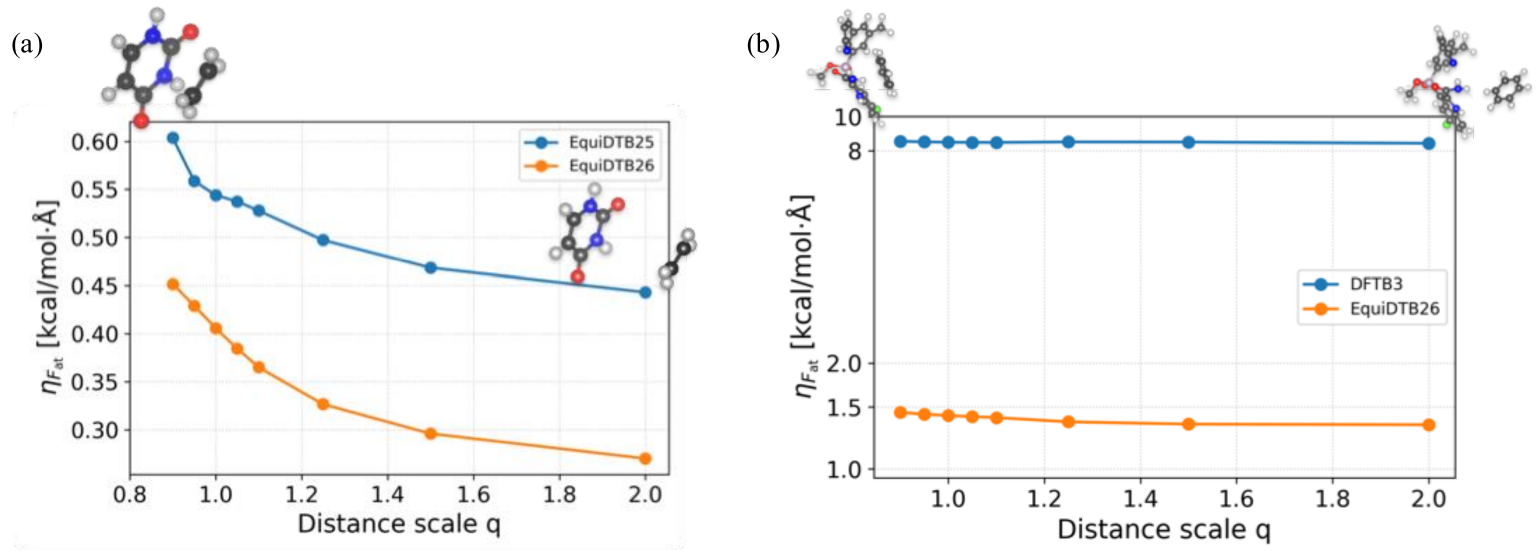}
    \caption{Variation of the mean absolute error (MAE) of atomic forces as a function of the intermolecular distance scaling factor $q$ for molecular dimers from the (a) S66$\times$8 and (b) QUID benchmarks. Results are shown for EquiDTB25 and EquiDTB26 using PBE0+MBD reference atomic forces. Error values represent averages over all dimers at each distance scale.    }
    \label{fig:dimers}
\end{figure}

\begin{figure}[t!]
    \centering
    \includegraphics[width=0.8\linewidth]{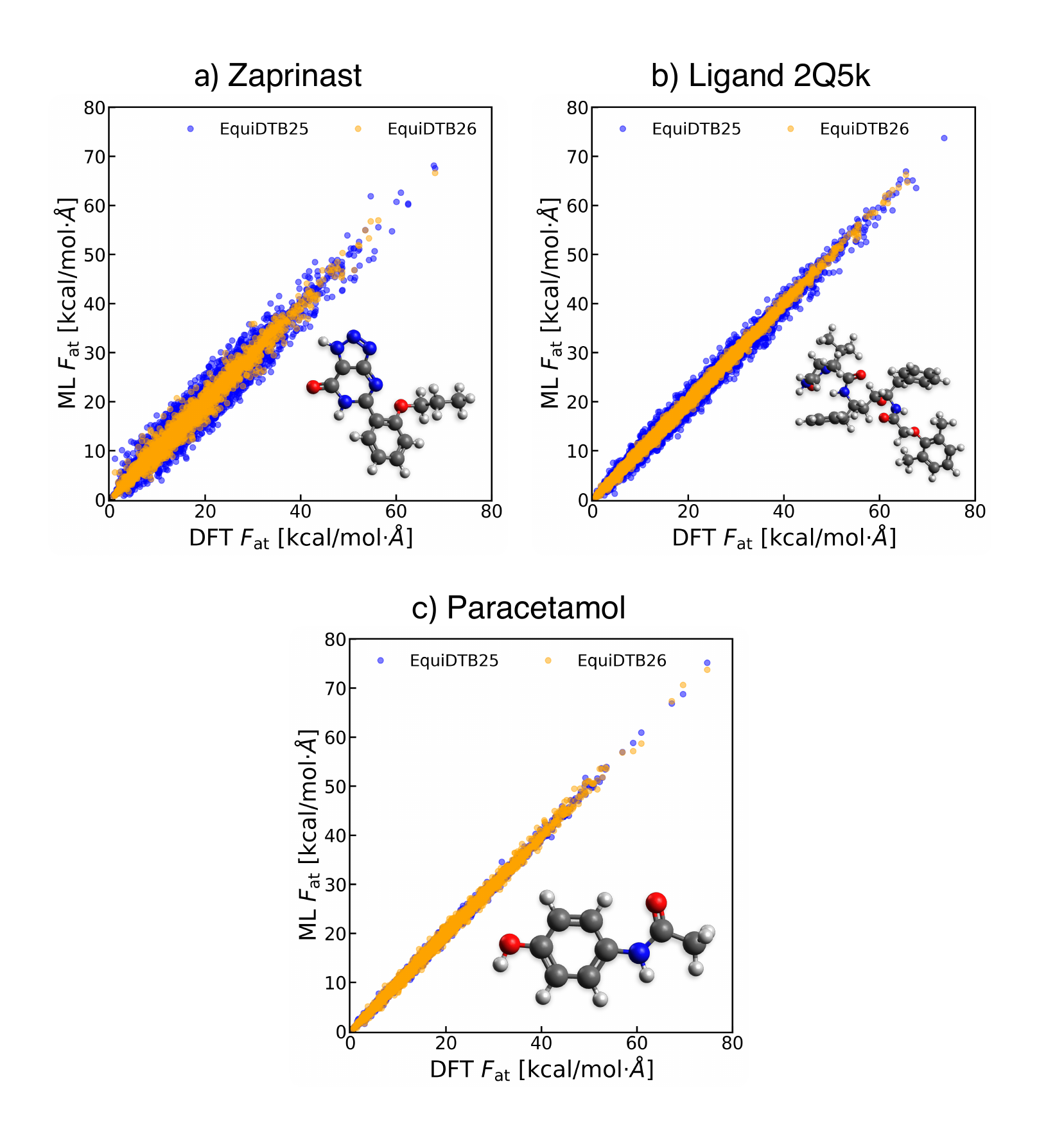}
    \caption{Correlation plots between PBE0+MBD reference atomic forces and those predicted by EquiDTB25 and EquiDTB26 for 200 molecular conformations extracted from molecular dynamics trajectories at 300 K of (a) paracetamol, (b) ligand 2Q5k, and (c) zaprinast. 
    }
    \label{fig:S1}
\end{figure}

To further assess the transferability of EquiDTB26 beyond the QCML test set, we performed validation calculations on three independent benchmark datasets probing different aspects of molecular-force prediction: the S66$\times$8 molecular dimers, flexible drug-like molecules, and the QUID benchmark. In all cases, performance was evaluated using the mean absolute error (MAE) of the atomic forces relative to PBE0+MBD reference calculations. The numerical results are summarized in Table~S\ref{tab:force_mae_models}, while representative correlations and benchmark trends are presented in Figs.~S\ref{fig:dimers},S\ref{fig:S1}.

\begin{table}[ht]
\centering
\caption{Mean absolute errors (MAEs) in the test set for energy and force predictions of models trained on different subsets of the QCML dataset. Energy errors are reported per atom.}
\label{tab:qcml_model_mae}
\begin{tabular}{lcc}
\hline
\textbf{Model} &
\textbf{Energy MAE} &
\textbf{Force MAE} \\
&
\textbf{(kcal/mol/atom)} &
\textbf{(kcal/mol/\AA)} \\
\hline
500k & 0.132 & 1.716 \\
1M   & 0.087 & 1.228 \\
2M   & 0.080 & 1.150 \\
3M   & 0.065 & 0.987 \\
4M (EquiDTB26)  & \textbf{0.059} & \textbf{0.905} \\
\hline
\end{tabular}
\end{table}

\paragraph{S66x8 molecular dimers.}

The robustness of the learned force correction was further assessed using the S66$\times$8 benchmark, which contains equilibrium and non-equilibrium molecular dimers sampled at different intermolecular separation distances. Following the procedure adopted in our previous work, the average force error was evaluated as a function of the relative distance scaling factor $q$, allowing model performance to be assessed across both compressed and elongated intermolecular geometries.
As shown in Fig.~S\ref{fig:dimers}(a), EquiDTB26 consistently exhibits lower force errors than EquiDTB25 across the full range of intermolecular distances. Averaged over all distance scales, the force MAE decreases from 0.522 to 0.366~kcal/mol/\AA{}, corresponding to an improvement of approximately 30\%. The reduction in error is observed for both compressed and elongated dimers, indicating that the expanded QCML training set enables EquiDTB26 to better capture force corrections associated with diverse non-covalent interactions while preserving the robustness of the original EquiDTB framework.

\paragraph{QUID benchmark.}

The performance of EquiDTB26 was also evaluated on the QUID benchmark, which comprises larger and chemically diverse ligands that extend well beyond the molecular sizes and chemical compositions represented in the original EquiDTB25 training dataset. Since QUID was not considered in our previous work, the comparison here is performed against the underlying DFTB3 method.
Fig.~S\ref{fig:dimers}(b)  shows the variation of the force MAE as a function of the molecular distance scaling factor $q$. While the force error obtained with DFTB3 remains nearly constant at approximately 8.5~kcal/mol/\AA{} across the entire range of structural distortions, EquiDTB26 reduces this value to approximately 1.4~kcal/mol/\AA{}, corresponding to an improvement of roughly a factor of six. The nearly constant error profile further indicates that the learned $\Delta_{\rm TB}$ correction remains stable not only near equilibrium geometries but also under substantial structural distortions.


\paragraph{Flexible molecules.}

The capability of EquiDTB26 to describe the potential-energy surfaces of larger molecules was assessed by analysing atomic forces extracted from molecular dynamics trajectories of paracetamol, ligand 2Q5k, and zaprinast. Fig.~S\ref{fig:S1} shows the correlation between reference PBE0 atomic forces and those predicted by EquiDTB25 and EquiDTB26 for 200 representative conformations of each molecule.
Overall, both models exhibit excellent linear correlations with the reference data. However, EquiDTB26 generally produces a tighter distribution around the ideal diagonal, indicating more accurate force predictions. This improvement is particularly evident for the larger and chemically more complex molecules. For ligand 2Q5k, the force MAE decreases from 0.433 to 0.329~kcal/mol/\AA{}, while for zaprinast it decreases from 1.059 to 0.717~kcal/mol/\AA{}. In contrast, both models perform comparably for paracetamol, with MAEs of 0.366 and 0.380~kcal/mol/\AA{} for EquiDTB25 and EquiDTB26, respectively. The comparable performance for paracetamol indicates that the broader chemical space introduced during training does not compromise the description of smaller molecules already well represented in the original training dataset. Conversely, the improvements observed for ligand 2Q5k and zaprinast demonstrate the enhanced transferability of EquiDTB26 to larger and chemically more diverse systems.

\begin{table}[ht]
\centering
\caption{Mean absolute error (MAE) of the atomic forces (kcal/mol/\AA{}) for the validation benchmarks obtained with EquiDTB25 and QCML-trained correction models of increasing training-set size. The best-performing model for each benchmark is highlighted in bold.}
\label{tab:force_mae_models}
\begin{tabular}{lccccc}
\hline
\textbf{Model} &
\multicolumn{3}{c}{\textbf{Flexible Molecules}} &
\textbf{S66$\times$8} &
\textbf{QUID} \\
\cline{2-4}
& \textbf{Zaprinast} & \textbf{Ligand 2Q5K} & \textbf{Paracetamol} & & \\
\hline
EquiDTB25 & 1.059 & 0.433 & 0.366 & 0.522 & -- \\
500k & 0.885 & 0.519 & 0.567 & 0.630 & 2.228 \\
1M & 0.711 & 0.399 & 0.453 & 0.432 & 1.588 \\
2M & 0.894 & 0.379 & \textbf{0.363} & 0.390 & 1.556 \\
3M & \textbf{0.700} & 0.349 & 0.393 & 0.436 & 1.484 \\
4M (EquiDTB26) & 0.717 & \textbf{0.329} & 0.380 & \textbf{0.366} & \textbf{1.409} \\
\hline
\end{tabular}
\end{table}

\clearpage

\section{AlloQM dataset}

\begin{table}[h]
    \centering 
    \caption{
    List of quantum-mechanical properties included in the alloQM dataset, which contains 6,253 conformations of 241 unique allosteric drug molecules.
    These features were calculated using the semi-empirical third-order DFTB method (DFTB3)  supplemented with a treatment of many-body dispersion (MBD) for van der Waals interactions. Structures were optimized using the EquiDTB26 model. 
    }
    \begin{tabular}{lll} \toprule
          Label & Property name & Dim \\ \hline 
         $\vec{R}$ & XYZ atomic coordinates & N,3 \\
         Z & Atomic numbers & N \\
         $\text{E}_\text{band}$ & Band energy & 1 \\
         $\text{E}_\text{H0}$ & Reference density energy  & 1 \\
         $\text{E}_\text{scc}$ & Self-consistent charge energy  & 1 \\
         $\text{E}_\text{3rd}$ & Third-order correction energy  & 1 \\
         $\text{E}_\text{rep}$ & Repulsion energy  & 1 \\
         $\text{E}_\text{mbd}$ & Many-body interaction energy  & 1 \\
         $\text{E}_\text{gap}$ & HOMO-LUMO energy gap  & 1 \\
         $\text{E}_\text{HOMO}$ & HOMO energy   & 1 \\
         $\text{E}_\text{LUMO}$ & LUMO energy   & 1 \\
         $\varepsilon$ & Molecular orbital energies & 8 \\ 
         $\mu$ & Scalar dipole moment & 1 \\ 
         $\vec{\mu}$ & Vector dipole moment & 3 \\ 
        $\vec{F}_{tot}$ & DFTB3+MBD atomic forces & N,3\\
         $Q$ & Atomic Mulliken charges & N \\ 
        \bottomrule
    \end{tabular}
    \label{tab:qmprops}
\end{table}

\begin{table*}[h]
  \centering
    \caption{List of navigation coordinates in the property space defined by the many-body dispersion energy $E_{\rm MBD}$ and HOMO--LUMO energy gap $E_{\rm gap}$ for QALPA model trained on the alloQM dataset. 
  }

    \begin{tabular}{lllllll}
    \hline\hline
   Property  & 1 & 2 & 3 & 4 & 5 & 6   \\
   \hline
  $N$ & 23 & 41 & 44 & 57 & 69 & 78   \\
      $E_{\rm MBD}$  & -0.5 & -1.0 & -1.4 & -1.9 & -2.2 & -2.5  \\
    $E_{\rm gap}$  & 4.0 & 3.1 & 2.8 & 2.7 & 3.2 & 3.5  \\

    \hline\hline 
  \end{tabular}
  \label{table7}
\end{table*}
